\documentclass[11pt]{article}

\usepackage{pb-diagram}
\usepackage{latexsym}
\usepackage{amsfonts}
\usepackage[usenames,dvipsnames]{xcolor}
\usepackage{graphicx,amssymb,amsmath,epsfig}
\usepackage[english]{babel}
\usepackage{graphicx}
\usepackage{dcolumn}
\usepackage{bm}

\definecolor{myblue}{rgb}{0,0,0.8}
\usepackage[colorlinks=true
,urlcolor=myblue	
,anchorcolor=myblue
,citecolor=myblue
,filecolor=myblue
,linkcolor=myblue
,menucolor=myblue
,pagecolor=myblue
,linktocpage=true  
]{hyperref}

\catcode`\@=11
\def\marginnote#1{}

\newcount\hour
\newcount\minute
\newtoks\amorpm
\hour=\time\divide\hour by60 \minute=\time{\multiply\hour by60
\global\advance\minute by-\hour}\edef\standardtime{{\ifnum\hour<12
\global\amorpm={am}%
        \else\global\amorpm={pm}\advance\hour by-12 \fi
        \ifnum\hour=0 \hour=12 \fi
        \number\hour:\ifnum\minute<10
0\fi\number\minute\the\amorpm}}
\edef\militarytime{\number\hour:\ifnum\minute<10 0\fi\number\minute}

\def\draftlabel#1{{\@bsphack\if@filesw {\let\thepage\relax
   \xdef\@gtempa{\write\@auxout{\string
      \newlabel{#1}{{\@currentlabel}{\thepage}}}}}\@gtempa
   \if@nobreak \ifvmode\nobreak\fi\fi\fi\@esphack}
        \gdef\@eqnlabel{#1}}
\def\@eqnlabel{}
\def\@vacuum{}
\def\draftmarginnote#1{\marginpar{\raggedright\scriptsize\tt#1}}
\def\draft{\oddsidemargin -.5truein
        \def\@oddfoot{\sl preliminary draft \hfil
        \rm\thepage\hfil\sl\today\quad\militarytime}
        \let\@evenfoot\@oddfoot \overfullrule 3pt
        \let\label=\draftlabel
        \let\marginnote=\draftmarginnote

\def\@eqnnum{(\theequation)\rlap{\kern\marginparsep\tt\@eqnlabel}%
\global\let\@eqnlabel\@vacuum}  }

\def\numberbysection{\@addtoreset{equation}{section}
        \def\theequation{\thesection.\arabic{equation}}}

\def\underline#1{\relax\ifmmode\@@underline#1\else
 $\@@underline{\hbox{#1}}$\relax\fi}

\catcode`@=12 \relax

\numberbysection

\topmargin by -\headsep
\def\nonu{\nonumber}
\def\br{\begin{eqnarray}}
\def\er{\end{eqnarray}}

\def\({\left(}
\def\){\right)}
\def\[{\left[}
\def\]{\right]}

\def\a{\alpha}

\def\bpsi{\bar{\psi}}

\def\g{\gamma}
\def\G{\Gamma}

\def\l{\lambda}

\def\o{\omega}

\def\pa{\partial}

\def\pt{\partial}

\def\s{\sigma}
\def\S{\Sigma}

\def\t{\tau}
\def\th{\theta}

\def\tp0{\Theta_{+}^{(0)}}
\def\tm0{\Theta_{-}^{(0)}}

\def\bp{{\bar \p}}

\def\cD{{\cal D}}

\def\bp{\bar{\psi}}
\def\bpp{\bar{\psi}_+}

\def\dx{\pt_x\phi}
\def\dxx{\pt^2_x\phi}
\def\dxxx{\pt^3_x\phi}
\def\dxxxx{\pt^4_x\phi}
\def\dxxxxx{\pt^5_x\phi}

\def\dxxxxxxx{\pt^7_x\phi}
\def\l{\lambda}
\def\th{\theta}
\def\nonu{\nonumber}
\def\ni{\noindent}
\def\bi{\begin{itemize}}
\def\ei{\end{itemize}}
\def\f{\phi}

\begin{document}

\vspace*{1cm}
\noindent

\vskip 1 cm
\begin{center}
{\Large\bf   Defects in the supersymmetric mKdV hierarchy via B\"acklund transformations}
\end{center}
\normalsize
\vskip 1cm
\begin{center}
{A.R. Aguirre}\footnote{\href{mailto:alexis.roaaguirre@unifei.edu.br}{alexis.roaaguirre@unifei.edu.br}}{$^\dagger$},  A.L. Retore\footnote{\href{mailto:retore@ift.unesp.br}{retore@ift.unesp.br}}{$^\ast$}, J.F. Gomes\footnote{\href{mailto:jfg@ift.unesp.br}{jfg@ift.unesp.br}}{$^\ast$}, N.I. Spano\footnote{\href{mailto:natyspano@ift.unesp.br}{natyspano@ift.unesp.br}}{$^\ast$}, and A.H. Zimerman\footnote{\href{mailto:zimerman@ift.unesp.br}{zimerman@ift.unesp.br}}{$^\ast$}\\[.7cm]

\par \vskip .1in \noindent
{$^\ast$} {Instituto de F\'isica Te\'orica - IFT/UNESP,\\
Rua Dr. Bento Teobaldo Ferraz, 271, Bloco II,
CEP 01140-070,\\ S\~ao Paulo - SP, Brasil.}\\[0.3cm]
{$^\dagger$} {Instituto de F\'isica e Qu\'imica, Universidade Federal de Itajub\'a - IFQ/UNIFEI,\\
Av. BPS 1303,  CEP 37500-903, Itajub\'a - MG, Brasil.}
\vskip 2cm

\end{center}

\begin{abstract}

The integrability  of the ${\cal N}=1$ supersymmetric modified Korteweg de-Vries (smKdV) hierarchy in the presence of defects  is investigated through the construction of 
its super B\"acklund transformation. The  construction of such transformation is performed by using essentially two methods: the B\"acklund-defect matrix approach
and the superfield approach. Firstly, we employ the defect matrix associated to the hierarchy which turns out  to be the same  for the supersymmetric sinh-Gordon (sshG) model.
The method is general  for all flows and   as an example  we  derive 
  explicitly 
the B\"acklund equations in components for the  first few flows  of the hierarchy, namely $t_3$ and $t_5$. 
Secondly,   the supersymmetric extension of the B\"acklund transformation 
in the superspace formalism is constructed for those flows. Finally, this super B\"acklund transformation is employed to introduce
 type I defects for 
the supersymmetric mKdV 
hierarchy.   Further   integrability aspects by considering modified conserved quantities  are derived from the defect matrix.

\end{abstract}

\newpage
\tableofcontents

\vskip 1cm
\hspace{-0.73cm}
\rule{16.5cm}{1pt}
\vskip 2.3cm

\section{Introduction}
\label{sec:intro}

The introduction of special impurities or defects within two-dimensional integrable models that  preserves the integrability properties has been 
recently an intensively studied topic. Integrable defects, as they are known, were introduced originally in \cite{Corr1,Corr2}, as a set of internal boundary 
conditions derived from a Lagrangian density located at certain spatial position connecting two types of solutions.

The presence of these special defects has been studied in several models, including sine(h)-Gordon \cite{Corr1}, affine Toda field theories \cite{Corr2}, 
the non-linear Schr\"odinger,  and other non-relativistic field theories \cite{jumpCorr, Cau}. An interesting feature shared by all of these defect systems is that the associated defect conditions correspond to frozen B\"acklund transformations \cite{Rogers}. This kind of defect is then named type-I if the fields
on either side of it only interact with each other at the defect location. It is called type-II if
they interact through additional degrees of freedom associated to the
defect itself which only exist at the defect point \cite{Corr09, Ale6}.

Several other interesting issues have been studied for these types of integrable defects, among which the following are worth mentioning: the computation of the higher order modified conserved
quantities and their involutivity \cite{Doi1,Doi2, Doi3}, quantum description \cite{Del1,Del2, Konik, Bowcock05, Zambon07, Kundu,Corrigan:2010sr,Corrigan:2010ph,Corr2011}  (see also \cite{Mikhail14, Mikhail15} for the discussion of soliton defects in quantum systems), the multisimplectic
description \cite{Cau2014,Caudrelier:2014gsa,Avan2015, Doikou2016}, finite-gap solutions \cite{Parini}, extensions for non-simply laced affine Toda models \cite{CR, CR14, Bristow16,Bristow17}, and 
fermionic \cite{Ale, Ale2, Ale4} and supersymmetric extensions \cite{FLZ1, FLZ2, Ale5, Aguirre,Nathaly}.

The main purpose of this paper is to propose an extension of the framework of integrable defects for the supersymmetric modified Korteweg-de Vries 
hierarchy through the construction of the associated super B\"acklund transformation, by using the Lax approach.  In refs. \cite{Aratyn, Ymai}, authors have 
shown that the smKdV and the super sinh-Gordon equations belong to the same integrable hierarchy based on the $\widehat{sl}(2,1)$ affine super Lie algebra. 
On the other hand, it was shown in ref. \cite{Ana_proceedings} that the spatial part of the bosonic B\"acklund transformation for the mKdV hierarchy is universal 
within the entire hierarchy.
 Using this fact, it was proposed in ref. \cite{Ana15} that the associated 
defect-gauge matrix is also universal  and provide the corresponding B\"acklund equations for the entire hierarchy. 
Therefore, it is quite natural to expect  that such  property  will be preserved for the  supersymmetric extension in the sense that  the super sinh-Gordon  and other  models within the hierarchy will share the same defect matrix.  

The presence of type I and type II integrable defects in the ${\cal N} = 1$ sshG model has been already investigated in \cite{FLZ1, Aguirre, Nathaly}, through 
the Lagrangian formalism and the Lax approach, where the associated modified conserved charges was derived directly from the corresponding defect matrices.  

The main goal of this paper is  the extension of the results obtained  in \cite{FLZ1, Aguirre, Nathaly} for the super sinh-Gordon to the entire  smKdV hierachy. 
This includes the  systematic construction of the different time components of the B\"acklund transformations and   modified conservation laws.  
Our results are derived   firstly from the  invariance  under gauge transformations of the algebraic zero curvature  representation  and secondly  
from the superfield formalism.

This paper is organized as follows. Section \ref{sec2} summarises the necessary ingredients to construct the supersymmetric mKdV hierarchy through the $ \widehat{\textit{sl}} $(2,1) superalgebra. 

In section \ref{sec3}, we derive the super B\"acklund transformation for the smKdV hierarchy using 
 the type-I defect matrix derived previously for the sshG model and  assumed to be universal within the hierarchy. 
 The key  observation is that the zero curvature  representation is invariant under  gauge transformation  connecting two different field configurations  of the same model. 
 This provides a general  framework  from where the   B\"acklund transformations for the various  flows can be derived.  Explicit examples are  worked out  for the first two 
 flows, namely $t_3$ and $t_5$.   Moreover   all results  are  also rewritten within the superfield formalism.
 
Section \ref{sec4} discusses the conservation laws.  Firstly  we introduce the formalism for a situation  in which the defect is absent  and generate   an infinite  number of conserved charges.  Those are  conserved with respect to all flows and in particular, we explicitly verify the conservation  of the simplest two charges with respect to the first two flows, namely $t_3$ and $t_5$.  Next, the introduction of the defect  is shown to require modification  of the charges  in order to ensure its conservation. This is  accomplished systematically   assuming the  defect matrix to be responsible for the transition  from one side to the other of the defect. Again explicit examples are given for $t_3$ and $t_5$.  

Section \ref{sec5} contains some final remarks and comments on future directions to  investigate. The explicit representation of the  $ \widehat{\textit{sl}} $(2,1)  superalgebra, and some technical computations as well as long expressions for the super B\"acklund transformations are contained 
in appendices \ref{appA} to \ref{appF}.


\section{The supersymmetric mKdV hierarchy}
\label{sec2}

In this section we present a brief review of the systematic construction of the supersymmetric mKdV hierarchy based on the affine Kac-Moody superalgebra  $\widehat{\cal G}=\widehat{\textit{sl}} $(2,1) \cite{Ymai}. 
 
Let us start by considering the  super Lie algebra ${\textit{sl}} $(2,1), which has four bosonic generators $\big\{ h_1, h_2, E_{\pm\alpha_1} \big\}$,  and four fermionic generators $ \big\{E_{\pm\alpha_2},E_{\pm(\alpha_1+\alpha_2)}\big\} $, where $\a_1$ is bosonic simple root and $\a_2$, $\a_1+\a_2$ are fermionic simple  roots. The affine $\widehat{\textit{sl}} $(2,1)  structure is introduced by extending each generator $T_a \in sl(2,1)$ to $T_a^{(n)}$,
where $ d $ is defined by $ \big[d,T_a^{(n)}\big]=nT_a^{(n)} $. The hierarchy is further specified by introducing a decomposition of the $\widehat{\textit{sl}} $(2,1) superalgebra through the definition of a constant grade one element $E^{(1)}$, where
\br
 E^{(2n+1)} = h_1^{(n+1/2)}+2h_2^{(n+1/2)} - E_{\a_1}^{(n)}-E_{-\a_1}^{(n+1)},
\er
and the so called principal grading operator
\begin{equation}
Q=2d+\frac{1}{2}h_1^{(0)}.
\end{equation} 
\noindent
The grading operator $Q$ and the constant element $E^{(1)}$ decompose the affine superalgebra $\widehat{\cal G} = \oplus \,\widehat{\cal G}_m = {\cal K} \oplus {\cal M}$, where 
$m$ is the degree of the subspace $\widehat{\cal G}_m$ according to $Q$, \mbox{$ \mathcal{K}=\big\{x\in \widehat{\mathcal{G}}\,/[\,x,E^{(1)}]=0 \big\}$} is the kernel of $E^{(1)}$, and ${\cal M}$ its complement, in the following way
\begin{align}
& \hat{\mathcal{G}}_{2n+1}=\left\{K_1^{(2n+1)},\,K_2^{(2n+1)}, \,M_1^{(2n+1)}\right\},\nonumber\\
& \hat{\mathcal{G}}_{2n}=\left\{M_2^{(2n)}\right\},\nonumber\\
& \hat{\mathcal{G}}_{2n+\frac{1}{2}}=\left\{F_2^{(2n+\frac{1}{2})},\,G_1^{(2n+\frac{1}{2})}\right\},\nonumber\\
& \hat{\mathcal{G}}_{2n+\frac{3}{2}}=\left\{F_1^{(2n+\frac{3}{2})},\,G_2^{(2n+\frac{3}{2})}\right\},
\label{grading}
\end{align}
where the generators $F_i, G_i, K_i$, and $M_i$ are defined as linear combinations of the $\widehat{\textit{sl}} $(2,1) generators \cite{Ymai}. The representation of these generators is given in appendix \ref{appA}.

Now, the construction of the integrable hierarchy is based on the zero curvature condition
\begin{equation}
\left[ \pt_x+A_{x},\pt_{t_N}+A_{t_N}\right]=0,
\end{equation}
\noindent where $ A_{t_N} $ and $ A_x $ are the Lax pair. In general,  $ A_x $ is defined as \mbox{$ A_x=E^{(1)}+A_0+A_{1/2}$}  where  $A_0+A_{1/2}\in {\cal M}$, i.e.,
\begin{equation}
A_0=uM_2^{(0)}, \qquad A_{1/2}=\sqrt{i}\bar{\psi}\,G_1^{(1/2)}. \label{eq2.5}
\end{equation}
\noindent 
Here, $ u $ and $ \bar{\psi} $ are the corresponding fields of the integrable hierarchy. Now, we will assume that $ A_{t_N}=D^{(N)}+D^{(N-1/2)}+...+D^{(1/2)}+D^{(0)} $ for 
the positive hierarchy, where $ D^{(m)} $ has grade $m$. Then, the equation to be solved reads
\begin{equation}
\left[ \pt_x+E^{(1)}+A_{0}+A_{1/2},\pt_{t_N}+D^{(N)}+D^{(N-1/2)}+...+D^{(1/2)}+D^{(0)}\right]=0.
\label{zerocurv}
\end{equation}
The solving method consists on splitting the above equation grade by grade, which leads us to the following set of relations,
\begin{align}
& \,\,(N+1): \hspace{0.5cm} \left[E^{(1)},D^{(N)}\right]=0,\nonumber\\
& \left(N+1/2\right):\hspace{0.2cm}  \left[E^{(1)},D^{(N-1/2)}\right] +\left[A_{1/2},D^{(N)}\right]=0, \nonumber\\
& \,\,(N):\hspace{1.4cm} \pt_xD^{N}+\left[A_0,D^{(N)}\right]+\left[E^{(1)},D^{(N-1)}\right]+\left[A_{1/2},D^{(N-1/2)}\right]=0,\nonumber\\
& \hspace{2.5cm}\vdots{} & \nonumber\\
& \, (1): \hspace{1.5cm} \pt_xD^{(1)}+\left[A_0, D^{(1)}\right]+\left[E^{(1)},D^{(0)}\right]+\left[A_{1/2},D^{(1/2)}\right]=0,\nonumber\\
& \left(1/2\right): \hspace{1.2cm} \pt_xD^{(1/2)}+\left[A_0, D^{(1/2)}\right]+\left[A_{1/2},D^{(0)}\right]-\pt_{t_N}A_{1/2}=0,\nonumber\\
& \,\,(0): \hspace{1.5cm} \pt_xD^{(0)}+\left[A_0, D^{(0)}\right]-\pt_{t_N}A_{0}=0.
\label{decomposition}
\end{align} 
Note that, the image part of the zero and the one-half grade components of (\ref{decomposition}) yields the time evolution for the fields introduced in eq. (\ref{eq2.5}). Now, it is possible to expand each term $D^{(m)}$ by using the generators in eq. \eqref{grading}, as follows%
\begin{align}
& D^{(2n+1)}={ \tilde{a}}_{2n+1}K_1^{(2n+1)}+\tilde{b}_{2n+1}K_2^{(2n+1)}+\tilde{c}_{2n+1}M_1^{(2n+1)}, \nonumber\\
& D^{(2n)}=\tilde{a}_{2n}M_2^{(2n)}, \nonumber\\
& D^{(2n+\frac{1}{2})}=\tilde{a}_{2n+\frac{1}{2}}F_2^{\left(2n+\frac{1}{2}\right)}+\tilde{b}_{2n+\frac{1}{2}}G_1^{\left(2n+\frac{1}{2}\right)}, \nonumber\\
& D^{(2n+\frac{3}{2})}=\tilde{a}_{2n+\frac{3}{2}}F_1^{\left(2n+\frac{3}{2}\right)}+\tilde{b}_{2n+\frac{3}{2}}G_2^{\left(2n+\frac{3}{2}\right)},\label{eq2.8}
\end{align}
\noindent 
where the $\tilde{a}_m, \tilde{b}_m$, and $\tilde{c}_m$ are functionals of the fields $u$ and $\bpsi$. 
Substituting this parameterization in eq. \eqref{decomposition}, one can solve recursively for all  $D^{(m)}, m=0,\cdots N$. 
Note that the Lax component $ A_x $ does not depend on the index $ N $, and  will be the same for the entire hierarchy.  It takes the following form (see for instance \cite{Aguirre, Aratyn}), 

\begin{equation}
\renewcommand{\arraystretch}{2}
A_x=\left(
\begin{array}{cc|c}
\lambda^{1/2}-\pt_x\phi & -1 & \sqrt{i}\,\bpsi \hspace{0.2cm}\\
-\lambda & \lambda^{1/2}+\pt_x\phi & \sqrt{i}\,\lambda^{1/2}\,\bpsi\\\hline
\sqrt{i}\,\lambda^{1/2}\,\bpsi &\sqrt{i}\, \bpsi & 2\,\lambda^{1/2}
\end{array}
\right),\label{Ax}
\end{equation}
where we have redefined $u=-\dx $, and $\l$ is the spectral parameter. This parametrization establishes the explicit relationship between the relativistic (sinh-Gordon) and non-relativistic (mKdV) field variables. 

In what follows we will apply the procedure and consider explicit  solutions of the integrable hierarchy equations (\ref{decomposition}) for the  simplest members. For $ N=3 $, we find that the solution for the Lax component  $A_{t_3} = D^{(3)} + D^{(5/2)} + D^{(2)} + D^{(3/2)} + D^{(1)} + D^{(1/2)} + D^{(0)}$, is given by
{\small
\begin{equation}
\renewcommand{\arraystretch}{2.1}
A_{t_3}=\left(
\begin{array}{cc|c}
a_0 + \l^{1/2} a_{1/2}  -\l \dx+ \l^{3/2}  & a_+ -\l & \mu_+ +\l^{1/2} \nu_+ +\l \sqrt{i}\bpsi\\[0.1cm]
-\l a_- -\l^2& -a_0 +\l^{1/2} a_{1/2} +\l \dx +\l^{3/2}& \l^{1/2}\mu_- +\l \nu_- +\l^{3/2} \sqrt{i}\bpsi\\\hline
\l^{1/2}\mu_- -\l \nu_- +\l^{3/2} \sqrt{i}\bpsi& \mu_+ - \l^{1/2} \nu_++\l \sqrt{i}\bpsi & 2\l^{1/2} a_{1/2} +2 \l^{3/2}
\end{array}
\right),\label{At3}
\end{equation}}
where 
\br
 a_0 &=&-\frac{1}{4} \left(\dxxx-2(\dx)^3+3i\dx\bp \pt_x\bp \right), \qquad a_{1/2} = -\frac{i}{2}\bp\pa_x\bp, \nonumber\\
 a_{\pm} &=& \frac{1}{2}\left(\dxx \pm (\dx)^2 \mp i\bp\pt_x\bp\right), \qquad \qquad \qquad \nu_\pm = \frac{\sqrt{i}}{2}\big(\pa_x\bp \pm\bp \pa_x\phi\big), \quad\mbox{}\\
 \mu_\pm &=& \frac{\sqrt{i}}{4}\big(\pt^2_x\bp \pm\dx\pt_x\bp \mp \bp\dxx-2\bp(\dx)^2\big).\nonumber
\er
The equations of motion, which correspond to the zero and one-half grade components of (\ref{decomposition}), are in this case the ${\cal N}=1$ supersymmetric mKdV equations, namely
\begin{align}
& 4\pt_{t_3}u=\pt^3_xu-6u^2\pt_xu+3i\bp\pt_x\left(u\pt_x\bp\right),\label{movu}\\
& 4\pt_{t_3}\bp=\pt^3_x\bp-3u\pt_x\left(u\bp\right).\label{movpsi}
\end{align}
Now, for the $N=5$ member, the solution  for the Lax component $A_{t_5} = D^{(5)} + D^{(9/2)} +\cdots + D^{(0)}$ is given explicitly in appendix \ref{appB}. In this case, we find the following equations of motion
\footnote{{It is known that for the bosonic mKdV hierarchy the equations of motion can be obtained from a recursion operator $R$ by $\frac{\pt u}{\pt t_{N+2}}=R\frac{\pt u}{\pt t_{N}}$, where $R=\frac{1}{4}{\mathbb D}^2-{\mathbb D}u{\mathbb D}^{-1}u=\frac{1}{4}{\mathbb D}^2-u^2-u_x{\mathbb D}^{-1}u$, ${\mathbb D}=\pt_x$ and ${\mathbb D}^{-1}$ its inverse \cite{Olver}. The supersymetric case deserves attention and will be investigated in further developments.}},
\br
16\pt_{t_5}u &=&\pt_x^{5}u-10(\pt_x u)^3-40 u(\pt_x u)(\pt_x^{2} u)-10 u^2(\pt_x^{3} u)+30 u^4 (\pt_x u)+5 i \pt_x\bp\pt_x(u\pt_x^2\bp)\nonu\\[0.2cm]
&&+5 i \bp\pt_x(u \pt_x^3\bp-4 u^3 \pt_x\bp+\pt_x u\pt_x^2\bp+\pt_x^2u \pt_x\bp),\label{t5u}\\[0.2cm]
16\pt_{t_5}\bp &=&\pt_x^5\bp-5 u\pt_x(u\pt_x^2\bp+2\pt_x u\pt_x\bp+\pt_x^2 u\bp)+10 u^2\pt_x(u^2\bp)-10(\pt_x u)\pt_x(\pt_x u\bp).\qquad \,\,\,\mbox{}\label{t5psi}
\er
It is worth pointing out that the negative integrable hierarchy can be also constructed by considering the following zero curvature condition,
\br
 \big[ \pa_x+E^{(1)}+A_0+A_{1/2}, \pa_{t_{-M}} + D^{(-M)} + D^{(-M+1/2)}+\cdots + D^{(-1)}+D^{(-1/2)}\big]=0.\qquad\mbox{}
\er
The solutions are in general non-local, however, for the simplest case of $N=-M=-1$, we find that the Lax component $A_{t_{-1}}=D^{(-1)} + D^{(-1/2)}$, corresponds to the ${\cal N}=1$ sshG equation \cite{Nathaly, Ymai}, i.e
\begin{equation}
\renewcommand{\arraystretch}{2}
A_{t_{-1}}=\left(
\begin{array}{cc|c}
\lambda^{-1/2} & -\l^{-1} \,e^{2\phi} & -\l^{1/2}\sqrt{i}\,{\psi} e^\phi\hspace{0.2cm}\\
-e^{-2\phi}& \lambda^{-1/2}& -\sqrt{i}\,{\psi} e^{-\phi}\\\hline
\sqrt{i}\,{\psi} e^{-\phi} & \l^{1/2} \sqrt{i}\, {\psi} e^{\phi}& 2\lambda^{-1/2}
\end{array}
\right).
\end{equation}
In this case, the fields $\phi$, $\bpsi$ and $\psi$ satisfy
\begin{subequations}
\br 
 \pa_{t_{-1}}\pa_x\phi &=& 2\sinh2\phi + 2 \bp\psi\sinh\phi,\\
 \pa_{t_{-1}}\bp&=& 2 \psi \cosh \phi, \label{eqsshg}\\
 \pa_x\psi &=& 2 \bp \cosh\phi,\label{eqparapsi}
\er
\end{subequations}
the equations of motion of the ${\cal N}=1$ sshG model in the light-cone coordinates $(x, t_{-1})$.
We note that the equations of motion for all members of the hierarchy are invariant under the following supersymmetric transformations,
\begin{align}
\delta\phi=\sqrt{i}\bar{\epsilon}\,\bp, \qquad \delta\bp=\frac{1}{\sqrt{i}}\bar{\epsilon}\,\dx, \label{susytransf}
\end{align}
where $\bar{\epsilon}$  is a Grassmannian parameter. The associated supercharge can be written as follows,
\br 
{\cal Q} = \int_{-\infty}^{\infty} dx \(\bp\dx + 2\psi \sinh\phi\), 
\er
where the field $\psi$ is defined by  eq.  \eqref{eqparapsi}


\section{Super-B\"acklund transformations}
\label{sec3}

In this section we derive a general method to  generate the super-B\"acklund transformations (sBT) for all  members of the hierarchy.  First of all, we will use the defect matrix associated to the hierarchy in order
to derive the sBT in components. Then, we use the superspace approach to express them in terms of superfields.  The key ingredient is the gauge invariance of the 
zero curvature representation  generated by the defect matrix which, in turn   is assumed to relate  two field configurations. As  explicit examples we 
construct the super-B\"acklund transformation for the first two flows, namely, 
  $N=3$ (smKdV) equation, and for  the $N=5$ super-equation.

\subsection{B\"acklund transformations from the defect matrix}
\label{sec3.1}

Based upon the fact that the spatial Lax operator is common to all members of the mKdV hierarchy, it has been shown recently that the spatial component  of 
the B\"acklund transformation, and consequently the associated defect matrix, are also common  and henceforth universal  within 
the entire hierarchy \cite{Ana_proceedings, Ana15}. Here, 
we will extend these results to the supersymmetric mKdV hierarchy starting from the defect matrix already derived for the ($N=-1$ member), the super sinh-Gordon equation. 
The so-called type-I defect matrix can be written as follows \cite{Aguirre},
\begin{equation}
\renewcommand{\arraystretch}{2}
	K=\left(
	\begin{array}{cc|c}
	\lambda^{1/2} & -\frac{2}{\omega^2}e^{\phi_+}\lambda^{-1/2} & -\frac{2\sqrt{i}}{\omega}e^{\frac{\phi_+}{2}}f_1 \hspace{0.2cm}\\
	-\frac{2}{\omega^2}e^{-\phi_+}\lambda^{1/2} & \lambda^{1/2} & -\frac{2\sqrt{i}}{\omega}e^{-\frac{\phi_+}{2}}f_1\lambda^{1/2}\\\hline
	\frac{2\sqrt{i}}{\omega}e^{-\frac{\phi_+}{2}}f_1\lambda^{1/2} & \frac{2\sqrt{i}}{\omega}e^{\frac{\phi_+}{2}}f_1 & \frac{2}{\omega^2}+\lambda^{1/2}
	\end{array}
	\label{Kmatrix}
	\right)
\end{equation}
\ni
where $ \phi_\pm =\phi_1\pm\phi_2$,  $\omega $ represents the B\"acklund parameter, and $f_1$ is an auxiliary fermionic field. The defect matrix  $K$, connecting two different configurations $\{\phi_1,\bp_1\}$ and $\{\phi_2,\bp_2\}$, satisfies the following gauge equation,
\begin{equation}
\pt_xK=KA_x(\phi_1,\bar{\psi}_1)-A_x(\phi_2,\bar{\psi}_2)K. \label{equ3.2}
\end{equation}

\ni
 Now, by substituting \eqref{Ax} and \eqref{Kmatrix} in eq. (\ref{equ3.2}), we get 
	
\begin{align}
	& \pt_x\phi_-=\frac{4}{\omega^2}\sinh(\phi_+)-\frac{2i}{\omega}\sinh(\frac{\phi_+}{2}) f_1\bpp,\label{dxphi-}\\
	& \bpsi_-= \frac{4}{\omega} \cosh\left(\frac{\phi_+}{2}\right)f_1,\\
	& \pt_xf_1=\frac{1}{\omega}\cosh(\frac{\phi_+}{2})\bpp\label{dxf1}.
\end{align}
the spatial part of the B\"acklund transformations, where we have denoted $\bp_\pm=\bp_1\pm\bp_2 $. To derive the time component of the transformation, 
we consider the corresponding temporal part of the Lax pair $A_{t_n}$. For the smKdV equation ($N=3$), the second gauge condition reads,
\begin{equation}
\pt_{t_3}K=KA_{t_3}(\phi_1,\bar{\psi}_1)-A_{t_3}(\phi_2,\bar{\psi}_2)K.
\end{equation}
By substituting \eqref{At3} and \eqref{Kmatrix} in the above  equation, we obtain
\br 
4\pt_{t_3}\phi_-&=&\frac{i}{\o}\left[\pt^2_x\phi_+\cosh\Big(\frac{\phi_+}{2}\Big)-\left(\pt_x\phi_+\right)^2\sinh\Big(\frac{\phi_+}{2}\Big)\]\bar{\psi}_+f_1 \nonu\\
 &&\!\!\!\! -\frac{i}{\o}\left[\pt_x\phi_+\cosh\Big(\frac{\phi_+}{2}\Big)\pt_x\bar{\psi}_+ - 2\sinh\Big(\frac{\phi_+}{2}\Big)\pt^2_x\bar{\psi}_+\right]f_1\nonu\\
&&\!\!\!\!   +\frac{2}{\o^2}\Big[2(\pt^2_x\phi_+)\cosh\phi_+ -\left(\pt_x\phi_+\right)^2\sinh\phi_+ +i\bar{\psi}_+(\pt_x\bar{\psi}_+)\sinh\phi_+\Big]\nonu\\
&& \!\!\!\! -\frac{96i}{\o^5}\left[\sinh\Big(\frac{\phi_+}{2}\Big)+4\sinh^3\Big(\frac{\phi_+}{2}\Big) +3\sinh^5\Big(\frac{\phi_+}{2}\Big)\right]\bar{\psi}_+f_1\nonu\\
&& \!\!\!\! -\frac{32}{\o^6}\,\sinh^3\phi_+,\label{dtphim}\\[0.1cm]
 4\pa_{t_3}f_1 &=& \frac{1}{2\o}\cosh\Big(\frac{\phi_+}{2}\Big) \left[2\pa_x^2\bp_+ -\bp_+ (\pa_x\phi_+)^2\] +\frac{1}{2\o}\sinh\Big(\frac{\phi_+}{2}\Big) \left[\bp_+\pa_x^2\phi_+ -\pa_x\phi_+\pa_x\bp_+\]\nonumber \\[0.1cm]
 &&-\frac{12}{\o^4}\sinh\phi_+\cosh^2\Big(\frac{\phi_+}{2}\Big) (\pa_x\phi_+)f_1 +\frac{12}{\o^5}\sinh^2\phi_+\cosh\Big(\frac{\phi_+}{2}\Big) \bp_+\label{dtf1}.
\er
Equations (\ref{dxphi-})--(\ref{dxf1}), and (\ref{dtphim}) and (\ref{dtf1}) correspond to the super-B\"acklund transformations 
for the smKdV in components. It can be easily verified that they are consistent by cross-differentiating any of them. Notice also that by
setting all the fermions to zero we recover the bosonic case, i.e.,  the B\"acklund transformation of the mKdV \cite{Ana_proceedings},
\begin{subequations}
\begin{align}
& \pt_x\phi_-=\frac{4}{\omega^2}\sinh\phi_+, \label{dxphibos}\\
& 4\pt_{t_3}\phi_-=\frac{4}{\o^2}\pt^2_x\phi_+\cosh\phi_+-\frac{2}{\o^2}\left(\pt_x\phi_+\right)^2\sinh\phi_+-\frac{32}{\o^6}\sinh^3\phi_+.\label{dt3phibos}
\end{align}
\end{subequations}

Now, to derive the temporal part of the super-B\"acklund transformation for the $N=5$ member of the hierarchy we consider the corresponding Lax operator $A_{t_5}$. 
 The gauge condition reads,
\br
 \pt_{t_5}K=KA_{t_5}(\phi_1,\bar{\psi}_1)-A_{t_5}(\phi_2,\bar{\psi}_2)K.
\er
By solving this condition for $A_{t_5}$ given in appendix \ref{appB}, we obtain
\begin{eqnarray}
	16\pt_{t_5}\phi_-& =& -\frac{i}{\omega}\left[c_0\, \bar{\psi}_+ + c_1\, \pt_x\bar{\psi}_+ +c_2\, \pt^2_x\bar{\psi}_+ +c_3\,\pt^3_x\bar{\psi}_++c_4\, \pt^4_x\bar{\psi}_+\right] f_1\nonumber\\
	&& +\frac{1}{\omega^2}\left[c_5+ ic_6\, \bar{\psi}_+\pt_x\bar{\psi}_++ ic_7\, \bar{\psi}_+\pt^2_x\bar{\psi}_++ic_8\big(\bar{\psi}_+\pt^3_x\bar{\psi}_+-(\pt_x\bar{\psi}_+)(\pt^2_x\bar{\psi}_+)\big)\right]\nonumber\\
	&& -\frac{i}{\omega^5}\left[\, c_9\,\bar{\psi}_++c_{10}\,\pt_x\bar{\psi}_++c_{11}\, \pt^2_x\bar{\psi}_+\, \right] f_1+\frac{1}{\omega^6}\left[\, c_{12}+i\, c_{13}\, \bar{\psi}_+\pt_x\bar{\psi}_+\, \right]\nonu\\
	&&+\frac{i}{\omega^9}\, c_{14} f_1\bar{\psi}_++\frac{c_{15}}{\omega^{10}},\label{btt5phi}\\[0.1cm]
	16\pt_{t_5}f_1 & =&\frac{1}{\omega}\left[\, g_0\, \bar{\psi}_++g_1\, \pt_x\bar{\psi}_+ +g_2\, \pt^2_x\bar{\psi}_+ +g_3\, \pt^3_x\bar{\psi}_+ +g_4\, \pt^4_x\bar{\psi}_+\right]\nonumber\\
	&& +\frac{1}{\omega^4}\left[g_6+ i\, g_5\,\bar{\psi}_+\pt_x\bar{\psi}_+\right]  f_1+\frac{1}{\omega^5}\left[\, g_7\, \bar{\psi}_++g_8\, \pt_x\bar{\psi}_++g_9\, \pt^2_x\bar{\psi}_+\, \right]\nonumber\\
	& &+\frac{g_{10}}{\o^8}\, f_1+\frac{g_{11}}{\o^9}\, \bar{\psi}_+,\label{btt5psi}
\end{eqnarray}
where $c_i,\, i=0,..,15$ and $g_j,\, j=0,..,11$ are functions depending on $\phi_+$ and its derivatives, and their explicit forms are given by 
(\ref{c0})-(\ref{c15}) and (\ref{d0})-(\ref{d11}) from appendix \ref{appC}, respectively. The equations (\ref{dxphi-})--(\ref{dxf1}), and (\ref{btt5phi}) and (\ref{btt5psi}) 
correspond to the super-B\"acklund transformations for the $N=5$ super equation.  Cross differentiating  (\ref{btt5phi}) and  (\ref{btt5psi})  
with respect of $x$ we recover the equations of motion (\ref{t5u}) and (\ref{t5psi}) after using equations.  (\ref{dxphi-})--(\ref{dxf1}).

Alternatively, for the bosonic case  the B\"acklund transformation for $t_5$ can be generated by  a B\"acklund recursion operator,
\begin{equation}\label{RBacklund}
\pt_{t_5}(\phi_1-\phi_2)=\frac{1}{4}\pt^2_x\pt_{t_{3}}(\phi_1-\phi_2)-\left(\pt_x\phi_1{\mathbb D}^{-1}\pt_x\phi_1{\mathbb D}\right)\pt_{t_{3}}\phi_1+\left(\pt_x\phi_2{\mathbb D}^{-1}\pt_x\phi_2{\mathbb D}\right)\pt_{t_{3}}\phi_2.
\end{equation}
Now, using that $\pt_{t_{3}}\phi_i=\frac{1}{4}\pt^3_x\phi_i-\frac{1}{2}(\pt_x\phi_i)^3, \, i=1,2$, we obtain in terms of $\phi_{\pm}=\phi_1\pm\phi_2$,
\begin{eqnarray}
\pt_{t_5}\phi_- &=&\frac{1}{16}\pt^5_x\phi_--\frac{5}{16}\pt^2_x\phi_-\pt^2_x\phi_+\pt_x\phi_+-\frac{5}{128}\pt_x\phi_-\left[4(\pt^2_x\phi_+)^2+8\pt_x\phi_+\pt^3_x\phi_+-3(\pt_x\phi_+)^4\right]\nonu\\
&&-\frac{5}{32}\pt^3_x\phi_-\left[(\pt_x\phi_+)^2+(\pt_x\phi_-)^2\right]+\frac{5}{64}\pt_x\phi_-\left[3(\pt_x\phi_-)^2(\pt_x\phi_+)^2-2\pt^2_x\phi_-\right]\nonumber\\&&+\frac{3}{128}(\pt_x\phi_-)^5.
\end{eqnarray}
Now, using the $x$-part of the B\"acklund (\ref{dxphibos})  we get,
\begin{eqnarray}
	16\pt_{t_5}\phi_-=\frac{c_5}{\omega^2}+\frac{c_{12}}{\omega^6}+\frac{c_{15}}{\omega^{10}},
\end{eqnarray}
where the functions $c_5, c_{12}, c_{15}$ are given in the appendix \ref{appC}. This equation corresponds precisely to the bosonic limit of (\ref{btt5phi}), when all fermions vanish. 
It is important to emphasize that for higher flows the  gauge condition,
 \br
 \pt_{t_N}K(\l)=K(\l)A_{t_N}(\phi_1,\bar{\psi}_1)-A_{t_N}(\phi_2,\bar{\psi}_2)K(\l).
\er
generates the respective B\"acklund transformation for the corresponding equation of motion.
It can be also shown that the B\"acklund equations (\ref{dxphi-})--(\ref{dxf1}), (\ref{dtphim}), (\ref{dtf1}), (\ref{btt5phi}) and (\ref{btt5psi}), are invariant under the supersymmetry transformations (\ref{susytransf}) if the auxiliary fermionic field $f_1$ transforms in the following way,
\begin{align}
\delta f_1=\frac{2\bar{\epsilon}}{\o\sqrt{i}}\sinh\Big(\frac{\phi_+}{2}\Big).
\end{align}

%

\subsection{Superspace formalism}
\label{super}

Let us now discuss the super B\"acklund transformations from the superfield point of view. We start by introducing
the fermionic superfield $\Psi(x,\th)=\sqrt{i} \bp(x) +\th u(x)$ to describe the supersymmetric extension of the mKdV equation \cite{Mathieu}, 
\br
 D_{t_3} \Psi = D^6\Psi - 3 (D\Psi)D^2(\Psi D\Psi),   \label{sfmkdv}
\er
where $\th$ is a Grassmannian coordinate,  $D=\pa_{\th} +\th \pa_x$ is the covariant super derivative, and we have defined  $D_{t_3}=4\pa_{t_3}$. In components, we recover equations (\ref{movu}) and (\ref{movpsi}), namely,
\br
4\pa_{t_3} u  &=& \pa_x^3 u - 6u^2\pa_x u +3i\bp \pa_x(u\pa_x\bp), \\
4\pa_{t_3} \bp &=& \pa_x^3 \bp - 3 u \pa_x(u\bp).
\er
Let us now define  a new bosonic superfield $\Phi(x,\th)=\phi(x)-\sqrt{i} \th \bp(x)$ \cite{Restuccia}, such that $\Psi = -D\Phi$, or equivalently $u=-\pa_x\phi$. Substituting in eq (\ref{sfmkdv}), we get 
\br 
 D_{t_3} \Phi = D^6\Phi - 2(D^2\Phi)^3 +3 (D\Phi)(D^2\Phi)(D^3\Phi). \label{DPhi}
\er
It is well-known that the spatial part of the super B\"acklund transformations for the hierarchy equations (\ref{dxphi-})--(\ref{dxf1}) can 
be derived from the following equations \cite{Kulish, Ymai09},
\br
  D\Phi_-&=& \frac{4i}{\omega}\cosh\Big(\frac{\Phi_+}{2}\Big)\S , \label{sb1}\\
  D\S &=& -\frac{2i}{\omega} \sinh\Big(\frac{\Phi_+}{2}\Big),\label{sb3}
\er
where $\Phi_\pm=\Phi_1\pm\Phi_2$, and $\S=-\frac{1}{\sqrt{i} } f_1+\th b_1$ is a fermionic superfield, with $f_1$ and $b_1$ being auxiliary fermionic and bosonic fields, respectively. In components, we find
\br
 \bp_- &=& \frac{4}{\o} \cosh\Big(\frac{\phi_+}{2}\Big)f_1,\\
 b_1 &=& -\frac{2i}{\omega}\sinh\Big(\frac{\phi_+}{2}\Big),\\
 \pt_x\phi_- &=&\frac{4}{\omega^2}\sinh\phi_+ +\frac{2i}{\omega}\sinh\Big(\frac{\phi_+}{2}\Big) \bp_+ f_1,\label{dxp}\\
\pt_x f_1 &=&\frac{1}{\omega}\cosh\Big(\frac{\phi_+}{2}\Big)\bp_+. \label{dxf}
\er
Now,  we propose the following supersymmetric extension of the ``temporal'' part of the super B\"acklund transformation for the smKdV ($N=3$) equation in the superspace,
\br
 D_{t_3}\Phi_-  &=& \frac{i}{\omega}\cosh\Big(\frac{\Phi_+}{2}\Big)\left[D^4\Phi_+ D\Phi_+-D^2\Phi_+D^3\Phi_+ \right]\S\nonumber\\
&&+\frac{i}{\o}\sinh\Big(\frac{\Phi_+}{2}\Big)\left[2 D^5\Phi_+-(D^2\Phi_+)^2(D\Phi_+) \]\S \label{sb2}\\
&&+\frac{2}{\o^2} \sinh\Phi_+\left[(D\Phi_+)(D^3\Phi_+)-(D^2\Phi_+)^2\]+\frac{4}{\o^2}\cosh\Phi_+ (D^4\Phi_+) \nonumber \\
&&-\frac{96i}{\o^5}\left[\sinh\Big(\frac{\Phi_+}{2}\Big)+4\sinh^3\Big(\frac{\Phi_+}{2}\Big)+3\sinh^5\Big(\frac{\Phi_+}{2}\Big)\](D\Phi_+)\S -\frac{32}{\o^6}\sinh^3\Phi_+, \qquad \mbox{}\nonumber \\[0.1cm]
 D_{t_3}\S &=& \frac{i}{2\o}\cosh\Big(\frac{\Phi_+}{2}\Big)\left[(D\Phi_+)(D^2\Phi_+)^2-2(D^5\Phi_+)\]\nonumber\\
 &&+\frac{i}{2\o}\sinh\Big(\frac{\Phi_+}{2}\Big)\left[(D^2\Phi_+)(D^3\Phi_+)-(D\Phi_+)(D^4\Phi_+)\]\nonumber \\[0.1cm]
 &&-\frac{12}{\o^4} \sinh\Phi_+\cosh^2\Big(\frac{\Phi_+}{2}\Big)(D^2\Phi_+)\S +\frac{12i}{\o^5}\sinh^2\Phi_+\cosh\Big(\frac{\Phi_+}{2}\Big) (D\Phi_+).\label{sb4}
\er
By cross-differentiating eqs. (\ref{sb1}) and (\ref{sb2}) we find that if $\Phi_1$ satisfies the smKdV equation (\ref{DPhi}), then $\Phi_2$ also satisfies it. Explicitly, in components we get
\br 
4\pt_{t_3}\phi_-&=&\frac{i}{\o}\left[\pt^2_x\phi_+\cosh\Big(\frac{\phi_+}{2}\Big)-\left(\pt_x\phi_+\right)^2\sinh\Big(\frac{\phi_+}{2}\Big)\]\bar{\psi}_+f_1 \nonu\\
 &&\!\!\!\! -\frac{i}{\o}\left[\pt_x\phi_+\cosh\Big(\frac{\phi_+}{2}\Big)\pt_x\bar{\psi}_+ - 2\sinh\Big(\frac{\phi_+}{2}\Big)\pt^2_x\bar{\psi}_+\right]f_1\label{equ3.26}\\
&&\!\!\!\!   +\frac{2}{\o^2}\Big[2(\pt^2_x\phi_+)\cosh\phi_+ -\left(\pt_x\phi_+\right)^2\sinh\phi_+ +i\bar{\psi}_+(\pt_x\bar{\psi}_+)\sinh\phi_+\Big]\nonu\\
&& \!\!\!\! -\frac{96i}{\o^5}\left[\sinh\Big(\frac{\phi_+}{2}\Big)+4\sinh^3\Big(\frac{\phi_+}{2}\Big) +3\sinh^5\Big(\frac{\phi_+}{2}\Big)\right]\bar{\psi}_+f_1  -\frac{32}{\o^6}\,\sinh^3\phi_+, \nonumber\\[0.1cm]
4\pt_{t_3} \bp_- &=& -\frac{1}{\o}\sinh\Big(\frac{\phi_+}{2}\Big)\left[(\pa_x\phi_+)^3-2 \pa_x^3\phi_+ -\frac{3}{2}(\pa_x\phi_+)(\bp_+\pa_x\bp_+)\] f_1\nonu\\
 &&+\frac{3}{\o^2}\sinh\phi_+ \left[\bp_+\pa^2_x\phi_+-(\pa_x\phi_+)(\pa_x\bp_+)\] \nonu\\
 && +\frac{2}{\o^2}\left[1+3\sinh^2\Big(\frac{\phi_+}{2}\Big)\]\left[2 \pa_x^2\bp_+ - \bp_+(\pa_x\phi_+)^2 \]\nonu \\[0.1cm]
&&-\frac{96}{\o^5}\left[\sinh\Big(\frac{\phi_+}{2}\Big)+4\sinh^3\Big(\frac{\phi_+}{2}\Big) +3\sinh^5\Big(\frac{\phi_+}{2}\Big)\right] (\pa_x\phi_+) f_1\nonu\\
&&+ \frac{24}{\o^6}\,\sinh^2\phi_+\big[1-7\cosh\phi_+\big]  \bp_+,\label{equ3.27}\\[0.1cm]
 4\pa_{t_3}f_1 &=& \frac{1}{2\o}\cosh\Big(\frac{\phi_+}{2}\Big) \left[2\pa_x^2\bp_+ -\bp_+ (\pa_x\phi_+)^2\] +\frac{1}{2\o}\sinh\Big(\frac{\phi_+}{2}\Big) \left[\bp_+\pa_x^2\phi_+ -\pa_x\phi_+\pa_x\bp_+\]\nonumber \\[0.1cm]
 &&-\frac{12}{\o^4}\sinh\phi_+\cosh^2\Big(\frac{\phi_+}{2}\Big) (\pa_x\phi_+)f_1 +\frac{12}{\o^5}\sinh^2\phi_+\cosh\Big(\frac{\phi_+}{2}\Big) \bp_+,\label{equ3.28}\\[0.1cm]
 4\pa_{t_3}b_1 &=& \frac{i}{2\o}\cosh\Big(\frac{\phi_+}{2}\Big) \left[(\pa_x\phi_+)^3-2 \pa_x^3\phi_+ -\frac{3i}{2}(\pa_x\phi_+)(\bp_+\pa_x\bp_+)\]\nonumber\\[0.1cm]
 &&+\frac{12}{\o^4}\sinh\phi_+\cosh^2\Big(\frac{\phi_+}{2}\Big)f_1\pa_x\bp_+ +\frac{6}{\o^4}\Big[\cosh\phi_+ +\cosh(2\phi_+)\Big]\bp_+f_1\pa_x\phi_+. \quad \mbox{}\label{equ3.29}
\er
We note that eqs. (\ref{equ3.27}) and (\ref{equ3.29}) can be derived from eqs. (\ref{equ3.26}) and (\ref{equ3.28}), and appear here only for consistency. 
It is also worth noting that eqs. (\ref{dxp}) and (\ref{dxf}) can be also derived from another set of B\"acklund equations by acting the super derivative operator on eqs. (\ref{sb1}) and (\ref{sb3}), namely
\br
 \pa_x\Phi_-&=& \frac{4}{\omega^2}\sinh\Phi_+ + \frac{2i}{\omega}\sinh\Big(\frac{\Phi_+}{2}\Big)(D\Phi_+)\S , \label{eq42}\\[0.1cm]
\pa_x\S &=&-\frac{i}{\omega}\cosh\Big(\frac{\Phi_+}{2}\Big)(D\Phi_+).
\er
The fermionic superfield $\S$ can be then eliminated from eq. (\ref{eq42}) by using eq. (\ref{sb1}) yielding the following relation,
\br
 \pa_x\Phi_- &=&\frac{4}{\omega^2}\sinh\Phi_+ + \frac{1}{2}\tanh\Big(\frac{\Phi_+}{2}\Big)(D\Phi_+)(D\Phi_-).\label{eq44BT}
\er
Analogously, from eq. (\ref{sb2}) we also have
\br
4\pa_{t_3} \Phi_- &=& \frac{2}{\o^2} \sinh\Phi_+\left[(D\Phi_+)(D^3\Phi_+)-(D^2\Phi_+)^2\]+\frac{4}{\o^2}\cosh\Phi_+ (D^4\Phi_+) -\frac{32}{\o^6}\sinh^3\Phi_+\nonumber \\[0.1cm]
&&-\frac{1}{4}\left[D^2\Phi_+D^3\Phi_+ -D^4\Phi_+ D\Phi_+\right](D\Phi_-),\nonumber\\[0.1cm]
&&-\frac{1}{4}\tanh\Big(\frac{\Phi_+}{2}\Big)\left[(D^2\Phi_+)^2(D\Phi_+) - 2 D^5\Phi_+\](D\Phi_-) \nonumber\\
&&-\frac{24}{\o^4}\tanh\Big(\frac{\Phi_+}{2}\Big)\left[1+4\sinh^2\Big(\frac{\Phi_+}{2}\Big)+3\sinh^4\Big(\frac{\Phi_+}{2}\Big)\](D\Phi_+)(D\Phi_-). \label{eq45BT}
\er
Equation (\ref{eq44BT}) has been previously obtained in \cite{Liu, Liu2, Xue2011}, and referred as a generalization of the B\"acklund transformation 
for the mKdV equation. However, that equation corresponds only to the spatial part of the transformation. Our method is systematic and generates also the time component of 
the  B\"acklund transformation for all flows.  
As an  example, eq. (\ref{eq45BT}) provides the 
temporal counterpart for $t_3$ which completes  a self consistent pair for  the B\"acklund transformation.

\subsection{Superextension for the $N=5$ equation}

Let us now construct the supersymmetric extension of the $N=5$ equation by considering the same fermionic superfield $\Psi(x,\th)=\sqrt{i} \bp(x) +\th u(x)$. To do that, we use the following direct extension procedure \cite{Mathieu, Lv},
\br
 16 u_{t_5} &\rightarrow& D_{t_5} \Psi, \nonu \\
 \pa_x^5 u &\rightarrow &D^{10} \Psi, \nonu\\
 (\pa_x u)^3 &\rightarrow& (D^2\Psi)(D^3\Psi)^2, \\
 u^2 (\pa_x^3u) &\rightarrow& \g_1 \Psi(D\Psi)(D^7\Psi) +(1-\g_1) (D\Psi)^2(D^6\Psi), \nonu\\
 u(\pa_x u)(\pa_x^2u) &\rightarrow& \g_2\Psi(D^3\Psi)(D^5\Psi) +\g_3(D\Psi)(D^2\Psi)(D^5\Psi)+(4-\g_2-\g_3)(D\Psi)(D^3\Psi)(D^4\Psi),\nonu\\
 u^4(\pa_xu) &\rightarrow & \g_4 (D^2\Psi) (D\Psi)^4 + (3-\g_4)\Psi (D\Psi)^3(D^3\Psi),\nonumber
\er
where we have introduced four free parameters $\g_k$. Then, the supersymmetric extension of the $N=5$ equation can be written by,
\br 
 D_{t_5} \Psi &=& D^{10}\Psi - 10 \Big[ (D^2\Psi)(D^3\Psi)^2 + \g_1 \Psi(D\Psi)(D^7\Psi) +(1-\g_1) (D\Psi)^2(D^6\Psi)\nonu\\
 && + \g_2\Psi(D^3\Psi)(D^5\Psi) +\g_3(D\Psi)(D^2\Psi)(D^5\Psi)+(4-\g_2-\g_3)(D\Psi)(D^3\Psi)(D^4\Psi)\nonu\\
 && - \g_4 (D^2\Psi) (D\Psi)^4 - (3-\g_4)\Psi (D\Psi)^3(D^3\Psi)\Big].
\er
In components, it takes the following form
\br 
16\pa_{t_5}u&=& \pa_x^5u -10\Big[(\pa_xu)^3+u^2(\pa_x^3 u) +4 u(\pa_xu)(\pa_x^2 u) - 3u^4 (\pa_x u) \Big]\nonu\\
&&+10\g_1\Big[iu\bpsi(\pa_x^4\bpsi)+ i(\pa_x^3u)\bpsi(\pa_x\bpsi)\Big] + 10\big(\g_3+\g_2-2\big)\Big[i(\pa_xu)(\pa_x\bpsi)(\pa_x^2\bpsi)\Big] \nonu \\
&& + 10\g_2\Big[i(\pa_xu)(\bpsi\pa_x^3\bpsi) + i(\pa_x^2u)(\bpsi\pa_x^2\bpsi)\Big]+10\big(\g_3 + 2\g_1 -2\big)\Big[iu(\pa_x\bpsi)(\pa_x^3\bpsi)\Big]\nonu \\
&&+10(\g_4-3)\Big[iu^3\bpsi(\pa_x^2\bpsi)+ 3iu^2(\pa_xu)(\bpsi\pa_x\bpsi)\Big],\label{e1st5}\\[01cm]
16\pa_{t_5}\bpsi&=& \pa_x^5 \bpsi -10 (\pa_x u)^2(\pa_x\bpsi) -10\g_1 u(\pa_x^3u) \bpsi +10(\g_1-1) u^2(\pa_x^3\bpsi) \nonu\\
&& -10\g_2 (\pa_x u)(\pa_x^2 u)\bpsi -10\g_3 \,u(\pa_x^2u)(\pa_x\bpsi) + 10\big(\g_2+\g_3-4\big) u(\pa_xu)(\pa_x^2\bpsi)\nonu\\
&& +10\g_4\, u^4 (\pa_x\bpsi) - 10(\g_4-3) u^3(\pa_xu)\bpsi.\label{e2st5}
\er
It can be noticed that eqs. (\ref{e1st5}) and (\ref{e2st5}) agree with eqs. (\ref{t5u}) and (\ref{t5psi}) when the parameters take the following values: $\g_1=1/2, \g_2=1, \g_3=3/2$, and $\g_4=1$. Therefore,  the supersymmetric extension of the $N=5$ equation reads as:
\br 
D_{t_5}\Psi &=& D^{10}\Psi - 5 \Big[ \Psi(D\Psi)(D^7\Psi)  + (D\Psi)^2(D^6\Psi) +2 (D^2\Psi) (D^3\Psi)^2 \nonu\\
&&+2\Psi(D^3\Psi)(D^5\Psi)  - 2(D\Psi)^4(D^2\Psi) +3(D\Psi)(D^2\Psi)(D^5\Psi) \nonu \\&& +3(D\Psi)(D^3\Psi)(D^4\Psi) -4\Psi(D\Psi)^3(D^3\Psi)\Big]. \label{SFt5}
\er
In terms of the bosonic superfield  $\Phi(x,\th) = \phi(x) - \sqrt{i} \th \bpsi(x)$,  the above equation takes the following form,
\br 
D_{t_5}\Phi &=& D^{10}\Phi + 5(D\Phi)(D^2\Phi)(D^7\Phi) +5(D\Phi)(D^3\Phi)(D^6\Phi) + 5(D\Phi)(D^4\Phi)(D^5\Phi)\nonu\\
&& -10(D^2\Phi)^2(D^6\Phi) - 10(D^2\Phi)(D^4\Phi)^2- 20(D\Phi)(D^2\Phi)^3(D^3\Phi)+6(D^2\Phi)^5.\qquad\mbox{}\label{Dt5Phi}
\er
Let us now construct the super B\"acklund transformation for the $N=5$ super equation. As we have already highlighted, the spatial part of the transformation is common to all members of the hierarchy, namely eqs. (\ref{sb1}) and (\ref{sb3}). Then, we propose the corresponding supersymmetric extension for the temporal part as follows,
\br
 D_{t_5} \Phi_- &=&  \frac{i \t_0}{\omega}\left[(D\Phi_+)(D^8\Phi_+)-(D^2\Phi_+)(D^7\Phi_+)-(D\Phi_+)(D^2\Phi_+)^2(D^4\Phi_+)\right. \nonumber \\  && \left. \qquad  \,\,\, +(D^2\Phi_+)^3(D^3\Phi_+) - (D^3\Phi_+)(D^6\Phi_+)+(D^4\Phi_+)(D^5\Phi_+)  \right] \Sigma\nonumber\\
 &&+ \frac{i\t_1}{\omega}\left[\frac{3}{4}(D^2\Phi_+)^4(D\Phi_+) -2 (D^2\Phi_+)^2(D^5\Phi_+) -3(D\Phi_+)(D^2\Phi_+)(D^6\Phi_+) \right. \nonumber \\&&\left.\qquad\quad  -4 (D^2\Phi_+)(D^3\Phi_+)(D^4\Phi_+)-(D\Phi_+)(D^4\Phi_+)^2+2(D^9\Phi_+)\right] \Sigma \nonumber\\
 &&+\frac{\t_2}{\omega^2}\left[2 (D\Phi_+)(D^7\Phi_+)-4(D^2\Phi_+)(D^6\Phi_+)-2(D^3\Phi_+)(D^5\Phi_+)+2(D^4\Phi_+)^2   \nonumber \right. \\&& \qquad\,\,\, \left. +\frac{3}{2} (D^2\Phi_+)^4 -4(D\Phi_+)(D^2\Phi_+)^2(D^3\Phi_+)\right]\nonumber \\
 &&+\frac{\t_3}{\omega^2} \left[2(D\Phi_+)(D^2\Phi_+)(D^5\Phi_+)+4(D\Phi_+)(D^3\Phi_+)(D^4\Phi_+) -6(D^2\Phi_+)^2(D^4\Phi_+)\right. \nonu \\ &&\qquad \,\,\, \left.+4(D^8\Phi_+)  \right] \nonumber\\ 
 && +\frac{i}{\omega^5} \left[\t_4(D\Phi_+) (D^2\Phi_+)^2 +\t_5 (D\Phi_+)(D^4\Phi_+)+\t_6 (D^2\Phi_+)(D^3\Phi_+)+\t_7 (D^5\Phi_+) \right]\Sigma\nonumber\\
 &&+\frac{1}{\omega^6}\left[\t_8(D\Phi_+)(D^3\Phi_+) + \t_9(D^2\Phi_+)^2 +\t_{10}(D^4\Phi_+) \right] \nonumber\\
 && +\frac{i\t_{11} }{\omega^9} (D\Phi_+) \Sigma + \frac{\t_{12}}{\omega^{10}},\label{equ3.40}
\er 
and
%
%
\br
 D_{t_5}\Sigma &=& \frac{i\s_0}{\omega} \left[3(D\Phi_+)(D^2\Phi_+)(D^6\Phi_+)-\frac{3}{4}(D^2\Phi_+)^4(D\Phi_+) +2 (D^2\Phi_+)^2(D^5\Phi_+) \right. \nonumber \\&&\left.\qquad  +4 (D^2\Phi_+)(D^3\Phi_+)(D^4\Phi_+)+(D\Phi_+)(D^4\Phi_+)^2-2(D^9\Phi_+)\right]\nonumber\\
 &&+\frac{i\s_1}{\omega}\left[(D\Phi_+)(D^8\Phi_+)-(D^2\Phi_+)(D^7\Phi_+)-(D\Phi_+)(D^2\Phi_+)^2(D^4\Phi_+)\right. \nonumber \\&&\left.\qquad\,\,\, +(D^2\Phi_+)^3(D^3\Phi_+)  - (D^3\Phi_+)(D^6\Phi_+)+(D^4\Phi_+)(D^5\Phi_+)  \right] \nonu \\
 &&+\frac{1}{\omega^4} \Big[ \s_2 (D\Phi_+)(D^2\Phi_+)(D^3\Phi_+) +\s_3 (D^2\Phi_+)^3 +\s_4 (D^2\Phi_+)(D^4\Phi_+)  +\s_5 (D^6\Phi_+)  \Big] \Sigma\nonumber\\
 && +\frac{i}{\omega^5} \Big[ \s_6 (D\Phi_+)(D^2\Phi_+)^2 + \s_7 (D\Phi_+)(D^4\Phi_+) + \s_8 (D^2\Phi_+)(D^3\Phi_+) +\s_9 (D^5\Phi_+) \Big] \nonumber\\
 && +\frac{\s_{10}}{\omega^8} (D^2\Phi_+)\Sigma 
 +\frac{i \s_{11}}{\omega^{9}}(D\Phi_+) ,\label{equ3.41}
\er
where the set of functions $\t_k$ and $\s_k$ depends on the superfield $\Phi_+$, and their explicit form are given in appendix \ref{appD}. After some algebra, it can be
verified that the super B\"acklund equations (\ref{btt5phi}) and (\ref{btt5psi}) are recovered from equations (\ref{equ3.40}) and (\ref{equ3.41}). 

It is worth pointing out that our method is  general and systematic for deriving the super B\"acklund transformation,  both from the defect matrix and  the superfield 
equation and   can be naturally extended to higher  flows within the  hierarchy. 
Here, we have illustrated how it works explicitly for the $N=3$ and $N=5$ members of the smKdV
hierarchy. Moreover, the same line of reasoning can be applied for any integrable hierarchy and its supersymmetric extension.


\section{Defects for the smKdV hierarchy}
\label{sec4}

In this section we will construct explicitly generating functions for an infinite set of independent conserved quantities for the smKdV hierarchy in the bulk theory 
and  derive the
corresponding modified conserved quantities arising from the defect contributions when a defect is present in the
theory by using the Lax approach.

\subsection{Conservation laws}\label{charges}
The associated linear problem for the smKdV hierarchy in the $(x, t_N)$ coordinates can be written as follows,
\begin{align}
& \pt_x\Omega(x,t_N;\l) = -A_x(x,t_N;\l)\Omega(x,t_N;\l)\label{problemx},\\
& \pt_{t_N}\Omega(x,t_N;\l) = -A_{t_N}(x,t_N;\l)\Omega(x,t_N;\l)\label{problemt}
\end{align}
where $\Omega=(\Omega_1,\Omega_2,\epsilon\,\Omega_3)^T$, with $\Omega_j$ bosonic components and $\epsilon$ is a fermionic parameter, and $\l$ is the spectral parameter. The compatibility of the above linear system yields the zero curvature equation,
\begin{equation}
\pt_{x}A_{t_N}-\pt_{t_N}A_x+\left[A_x,A_{t_N}\right]=0.
\end{equation}
Now, in order to construct a generating function for the conservation laws, we define the auxiliary functions $\G_{21}=\Omega_2\Omega_1^{-1}$ and $\G_{31}=\epsilon\Omega_3\Omega_1^{-1}$. Then, by considering the auxiliary problem (\ref{problemx}) and (\ref{problemt}), we find the following  conservation equation,
\begin{align}\label{conservation}
\pt_{t_N}\Big[V_{11}+V_{12}\G_{21}+V_{13}\G_{31}\Big]=\pt_x\Big[U_{11}+U_{12}\G_{21}+U_{13}\G_{31}\Big],
\end{align}  
where we have redefined $V=-A_x$ and $U=-A_{t_N}$ for simplicity, and  the functions 
$\G_{21}$ and $\G_{31}$ satisfy the following Ricatti equations, 
\begin{align}
&\pt_x\G_{21} = V_{21}+(V_{22}-V_{11})\G_{21}-V_{12}(\G_{21})^2+V_{23}\G_{31}-V_{13}\G_{31}\G_{21},\label{equ4.5}\\
&\pt_{x}\G_{31} = V_{31}+(V_{33}-V_{11})\G_{31}+V_{32}\G_{21}-V_{12}\G_{21}\G_{31},\label{equ4.6}\\
&\pt_{t_N}\G_{21} = U_{21}+(U_{22}-U_{11})\G_{21}-U_{12}(\G_{21})^2+U_{23}\G_{31}-U_{13}\G_{31}\G_{21}\label{eq4.7},\\
&\pt_{t_N}\G_{31} = U_{31}+(U_{33}-U_{11})\G_{31}+U_{32}\G_{21}-U_{12}\G_{21}\G_{31}\label{eq4.8}.
\end{align} 
Therefore, the corresponding first generating function of the conserved charges is given by, 
\begin{equation}
I_1=\int_{-\infty}^{\infty}dx\left[V_{11}+V_{12}\G_{21}+V_{13}\G_{31}\right]=\int_{-\infty}^{\infty}dx\left[-\l^{1/2}+\dx+\G_{21}-\sqrt{i}\bp\G_{31}\right]. \label{equ4.9}
\end{equation}
In order to get the explicit form for the conserved quantities, we consider the expansion of  $\G_{21}$ and $\G_{31}$ in powers of the spectral parameter $\l$ in such way that we can solve recursively the Riccati equations.
Let us expand $\G_{21}$ and $\G_{31}$ as $\l\rightarrow\infty$
\begin{equation}
\G_{21}=\sum_{n=-1}^{\infty}\l^{-n/2}\, \G_{21}^{(n/2)},\qquad \G_{31}=\sum_{n=0}^{\infty}\l^{-n/2}\, \G_{31}^{(n/2)}.
\end{equation}
By substituting these expansions into the Riccati equations (\ref{equ4.5}) and (\ref{equ4.6}) we find that the
first coefficients are given by,
\br
\G_{21}^{(-1/2)}&=&1,\qquad \G_{21}^{(0)}=-\dx,\qquad \G_{21}^{(1/2)}=\frac{1}{2}\dxx+\frac{1}{2}(\dx)^2,\qquad \G_{31}^{(0)}=-\sqrt{i}\bp, \\
\G_{21}^{(1)}&=&-\frac{1}{2}\dx\,\dxx-\frac{1}{4}\dxxx-\frac{i}{4}\bp\,\pt_x\bp\,\dx,\qquad \quad \G_{31}^{(1/2)}=\frac{\sqrt{i}}{2}(\pt_x\bp+\bp\,\dx),\\[0.1cm]
\G_{21}^{(3/2)}&=&\frac{1}{8}(\dxx)^2+\frac{1}{8}\dxxxx+\frac{1}{4}\dxxx\,\dx-\frac{1}{4}\dxx\,(\dx)^2-\frac{1}{8}(\dx)^4+\frac{i}{4}\bp\,\pt^2_x \bp\,\dx
\nonu\\
&&+\frac{i}{4}\bp\,\pt_x \bp\,\dxx+\frac{i}{4}\bp\,\pt_x \bp\,(\dx)^2,\\
\G_{31}^{(1)}&=&-\frac{\sqrt{i}}{4}(\pt^2_x\bp+\pt_x\bp\,\dx+\bp\,\dxx),\\
\G_{21}^{(2)}&=&-\frac{1}{8}\dxx\,\dxxx-\frac{1}{16}\dxxxxx+\frac{1}{2}(\dxx)^2\,\dx-\frac{1}{8}\dxxxx\,\dx+\frac{1}{4}(\dx)^2\dxxx\nonu\\
&&+\frac{1}{2}(\dx)^3\,\dxx-\frac{5i}{16}\bp\pt^2_x\bp\dxx-\frac{3i}{16}\bp\pt_x\bp\dxxx-\frac{3i}{16}\bp\pt^3_x\bp\dx\nonu\mbox{}\\
&& -\frac{i}{8}\pt_x\bp\pt^2_x\bp\dx-\frac{3i}{8}\bp\pt_x\bp\dxx \dx+\frac{i}{8}\bp\,\pt_x\bp\,(\dx)^3-\frac{i}{4}\bp\,\pt^2_x\bp\,(\dx)^2,\\
\G_{31}^{(3/2)}&=&\frac{\sqrt{i}}{8}\left[\pt^3_x\bp+\pt^2_x\bp\dx+\pt_x\bp\dxx-\pt_x\bp(\dx)^2+\bp\dxxx-\bp\dxx\dx-\bp(\dx)^3\right]\!\!.\qquad\,\,\,\mbox{}
\er
By substituting the coefficients of the expansion of the auxiliary functions in eq. (\ref{equ4.9}), we obtain the lowest non-trivial conserved charges $I_1^{(-n/2)}$, namely,
\br
I_1^{(-1/2)}&=&\frac{1}{2}\int_{-\infty}^{\infty}dx\left[\dxx+(\dx)^2-i\bp\,\pt_x\bp\right],\\
I_{1}^{(-1)}&=&-\frac{1}{4}\int_{-\infty}^{\infty}dx\left[2\dx\dxx+\dxxx-i\bp\,\pt^2_x\bp\right],\\
I_{1}^{(-3/2)}&=&\frac{1}{8}\int_{-\infty}^{\infty}dx\left[(\dxx)^2+\dxxxx+2\dxxx\dx-2\dxx(\dx)^2-(\dx)^4+i\bp\,\pt_x\bp\,\dxx\right.\nonu\\
&&\left.  \qquad \qquad \,\, +3i\bp\,\pt_x\bp\,(\dx)^2+i\bp\,\pt^2_x\bp\,\dx-i\bp\,\pt^3_x\bp\right].
\er
Analogously, we can construct a second set of conserved quantities from the second conservation law,
\begin{align}
\pt_{t_N}\Big[V_{22}+V_{21}\G_{12}+V_{23}\G_{32}\Big]=\pt_x\Big[U_{22}+U_{21}\G_{12}+U_{23}\G_{32}\Big],
\end{align}  
where the auxiliary fields $\G_{12}=\Omega_1\Omega_2^{-1}$ and $\G_{32}=\epsilon\Omega_3\Omega_2^{-1}$ now satisfy the Ricatti equations 
\begin{align}
&\pt_x\G_{12} = V_{12}-(V_{22}-V_{11})\G_{12}-V_{21}(\G_{12})^2+V_{13}\G_{32}-V_{23}\G_{32}\G_{12},\\
&\pt_x\G_{32} =V_{32}+(V_{33}-V_{22})\G_{32}+V_{31}\G_{12}-V_{21}\G_{12}\G_{32},\\
&\pt_{t_N}\G_{12} = U_{12}-(U_{22}-U_{11})\G_{12}-U_{21}(\G_{12})^2+U_{13}\G_{32}-U_{23}\G_{32}\G_{12},\\
&\pt_{t_N}\G_{32} = U_{32}+(U_{33}-U_{22})\G_{32}+U_{31}\G_{12}-U_{21}\G_{12}\G_{32} .
\end{align}
Now, the second generating function of conserved charges reads,
\begin{equation}
I_2=\int_{-\infty}^{\infty}\left[V_{22}+V_{21}\G_{12}+V_{23}\G_{32}\right]=\int_{-\infty}^{\infty}\left[-\l^{1/2}-\dx+\l\G_{12}-\sqrt{i}\l^{1/2}\bp\G_{32}\right].
\end{equation}
By using the expansions of $\G_{12}$ and $\G_{32}$ as $\l\rightarrow\infty$
\begin{equation}
\G_{12}=\sum_{n=1}^{\infty}\l^{-n/2}\, \G_{12}^{(n/2)},\qquad\G_{32}=\sum_{n=0}^{\infty}\l^{-n/2}\, \G_{32}^{(n/2)},
\end{equation}
we find
\br
\G_{12}^{(1/2)}&=&1,\qquad\G_{12}^{(1)}=\dx,\qquad \G_{32}^{(0)}=0,\qquad\G_{32}^{(1/2)}=-\sqrt{i}\bp,\\
\G_{12}^{(3/2)}&=&-\frac{1}{2}\dxx+\frac{1}{2}(\dx)^2,\qquad\G_{12}^{(2)}=\frac{1}{4}\dxxx-\frac{1}{2}\dx\dxx+\frac{i}{4}\bp\,\pt_x\bp\,\dx,\\
\G_{32}^{(1)}&=&\frac{\sqrt{i}}{2}(\pt_x\bp-\bp\,\dx),\qquad\G_{32}^{(3/2)}=\frac{\sqrt{i}}{4}(\bp\,\dxx+\pt_x\bp\,\dx-\pt^2_x\bp),\\
\G_{12}^{(5/2)}&=&\frac{1}{8}(\dxx)^2-\frac{1}{8}\dxxxx+\frac{1}{4}\dxxx\,\dx+\frac{1}{4}\dxx\,(\dx)^2-\frac{1}{8}(\dx)^4-\frac{i}{4}\bp\,\pt_x\bp\,\dxx\nonu\\
&&+\frac{i}{4}\bp\,\pt_x\bp\,(\dx)^2-\frac{i}{4}\bp\,\pt^2_x\bp\,\dx,\\
\G_{32}^{(2)}&=&\frac{\sqrt{i}}{8}\left[\pt^3_x\bp-\pt^2_x\bp\dx-\pt_x\bp\dxx-\pt_x\bp(\dx)^2-\bp\dxxx-\bp\dx\dxx+\bp(\dx)^3\right].\qquad \,\,\mbox{}
\er
Then, by substituting the above coefficients for the expansion of the auxiliary functions, we get the second set of non-trivial conserved quantities $I_2^{(-n/2)}$, 
\br
I_2^{(-1/2)}&=&-\frac{1}{2}\int_{-\infty}^{\infty}dx\left[\dxx-(\dx)^2+i\bp\,\pt_x\bp\right],\\
I_{2}^{(-1)}&=&-\frac{1}{4}\int_{-\infty}^{\infty}dx\left[2\dx\,\dxx-\dxxx-i\bp\,\pt^2_x\bp\right],\\
I_{2}^{(-3/2)}&=&\frac{1}{8}\int_{-\infty}^{\infty}dx\left[(\dxx)^2-\dxxxx+2\dxxx\dx+2\dxx(\dx)^2-(\dx)^4-i\bp\,\pt_x\bp\,\dxx\right.\nonu\\
&& \left. \qquad \qquad \,\,+3i\bp\,\pt_x\bp\,(\dx)^2-i\bp\,\pt^2_x\bp\,\dx-i\bp\,\pt^3_x\bp\right].
\er
The canonical energy and momentum in the bulk theory for $N=3$ member of the hierarchy are then recovered by simple combinations of the conserved quantities derived above, namely
\br
P&=&I_1^{(-1/2)}+I_2^{(-1/2)}=\int_{-\infty}^{\infty}dx\left[(\dx)^2-i\bp\,\pt_x\bp\right],\label{Pbulk}\er
and
\br
E&=&I_1^{(-3/2)}+I_2^{(-3/2)}\nonumber\\&=&\frac{1}{4}\int_{-\infty}^{\infty}dx\left[(\dxx)^2-(\dx)^4+2\dxxx\,\dx-i\bp\,\pt_x^3\bp+3i(\dx)^2\bp\pt_x\bp\right].\label{Ebulk}\qquad \mbox{}
\er 
Note that if the fermions vanish we recover the canonical momentum and energy for the mKdV \cite{jumpCorr}.
It is important to note that this conservation is valid for the entire hierarchy, since we have constructed the conserved charges for the smKdV hierarchy using only the spatial
part of the Lax operator, which is common for all members of the hierarchy. In fact, by taking the $N=3$ and $N=5$ time derivatives of the charge (\ref{Pbulk}), and using the equations of motion (\ref{movu}), (\ref{movpsi}), (\ref{t5u}), and (\ref{t5psi}), we get the following surface terms
\br\label{dtP}
\frac{dP}{dt_3}= \left[\frac{1}{2}\dx\dxxx-\frac{3}{4}(\dx)^4-\frac{1}{4}(\dxx)^2+\frac{9i}{4}(\dx)^2\bp\pt_x\bp+\frac{i}{2}\pt_x\bp\pt_x^2\bp-\frac{i}{4}\bp\pt_x^3\bp \right]_{-\infty}^{+\infty},\quad\,\,	\mbox{}
\er
and
\br
\frac{dP}{dt_5}&=&\!\left[\frac{5}{8}(\dx)^6+\frac{1}{16}(\dxxx)^2-\frac{1}{8}\dxx\,\dxxxx+\frac{1}{8}\dx\,\dxxxxx-\frac{5}{8}(\dx)^2(\dxx)^2-\frac{5}{4}(\dx)^3\dxxx\right.\qquad\,\, \mbox{}\nonumber\\
&&\left. \,\,+i\bp\pt_x\bp\left(\frac{35}{16}\dx\,\dxxx+\frac{5}{8}(\dxx)^2-\frac{25}{8}(\dx)^4\right)+\frac{15i}{16}\dx\,\dxx\bp\pt^2_x\bp\right.\nonumber\\
&&\left. \,\, +\frac{15i}{16}(\dx)^2\bp\pt^3_x\bp- \frac{5i}{8}(\dx)^2\pt_x\bp\pt^2_x\bp-\frac{i}{16}\bp\pt^5_x\bp+\frac{i}{8}\pt_x\bp\pt^4_x\bp-\frac{i}{8}\pt^2_x\bp\pt^3_x\bp\right]_{-\infty}^{+\infty}\!\!\!,
\er
which vanish since we are considering sufficiently smooth decaying
fields at $\pm\infty$. Analogously, for the canonical energy (\ref{Ebulk}),{\footnote{ Notice that  although the charges (\ref{Pbulk}) and (\ref{Ebulk}) are conserved with respect to all  flows, they are interpreted as canonical momentum and canonical energy, respectively, only for the $t_3$ model.}}  taking the $t_3$ and $t_5$ derivatives and using the equations of motion (\ref{movu}), (\ref{movpsi}), (\ref{t5u}), and (\ref{t5psi}), we obtain 
\br\label{dtE}
\frac{dE}{dt_3}&=&\left[\, \frac{1}{8}\dx\,\dxxxxx+\frac{1}{4}(\dx)^6+\frac{1}{16}(\dxxx)^2-\frac{3}{2}(\dx)^2(\dxx)^2-(\dx)^3\dxxx\right.\nonumber\\
&&\left. \,\,-\frac{3i}{16}\left((\dx)^4-5(\dxx)^2-9\dx\,\dxxx\,\right)\bp\pt_x\bp+\frac{3i}{2}\dx\,\dxx\,\bp\pt^2_x\bp+\frac{3i}{4}(\dx)^2\,\bp\pt^3_x\bp\right.\nonumber\\
&&\left.\,\,-\frac{i}{16}\bp\pt^5_x\bp+\frac{i}{16}\pt_x\bp\pt^4_x\bp-\frac{i}{16}\pt^2_x\bp\pt^3_x\bp\, \right]_{-\infty}^{+\infty},
\er
and
\br
\frac{dE}{dt_5}&=&\left[\, \frac{1}{64}(\dxx)^4-\frac{15}{64}(\dx)^8-\frac{1}{64}(\dxxxx)^2-\frac{3}{2}(\dx)^2(\dxxx)^2+\frac{1}{32}\dx\,\dxxxxxxx-\frac{3}{8}(\dx)^3\dxxxxx\right.\nonumber\\
&&+\left.
\frac{125}{32}(\dx)^4(\dxx)^2+\frac{25}{16}(\dx)^5\dxxx+\frac{1}{32}\dxxx\,\dxxxxx-\frac{41}{16}\dx(\dxx)^2\dxxx\right.\nonumber\\
&&-\frac{27}{16}(\dx)^2\dxx\,\dxxxx +\left.
i\bp\pt^2_x\bp\Big(\frac{69}{64}\dx\,\dxxxx-\frac{335}{64}(\dx)^3\dxx+\frac{127}{64}\dxx\,\dxxx\Big)\right.\nonumber\\
&&\left. +\frac{59i}{64}\dx\,\dxx\bp\pt^4_x\bp +
i\bp\pt_x\bp\left(\frac{23}{32}\dx\,\dxxxxx+\frac{55}{32}(\dx)^6-\frac{285}{32}(\dx)^2(\dxx)^2\right.\right.\nonumber\\
&&\left.\left. -\frac{375}{64}(\dx)^3\dxxx+\frac{89}{64}\dxx\,\dxxxx+\frac{73}{64}(\dxxx)^2\right) +
 i\bp\pt^3_x\bp\left(\frac{111}{64}\dx\,\dxxx-\frac{85}{64}(\dx)^4 \right. \right.\nonumber\\
 &&\left. \left. +\frac{7}{8}(\dxx)^2\right)-\frac{i}{64}\bp\pt^7_x\bp+\frac{i}{64}\pt_x\bp\pt^6_x\bp+
 \frac{33i}{64}\dx\,\dxx\,\pt_x\bp\pt^3_x\bp-\frac{i}{64}\pt^2_x\bp\pt^5_x\bp\right.\nonumber\\
 &&\left. -i\pt_x\bp\pt^2_x\bp\left(\frac{9}{64}\dx\,\dxxx+\frac{5}{32}(\dx)^4+\frac{39}{64}(\dxx)^2\right) +\frac{21i}{64}(\dx)^2\pt^2_x\bp\pt^3_x\bp \right. \nonumber\\ &&\left.  +\frac{9i}{64}(\dx)^2\pt_x\bp\pt^4_x\bp +\frac{i}{32}\pt^3_x\bp\pt^4_x\bp+\frac{9i}{32}(\dx)^2\bp\pt^5_x\bp\, \right]_{-\infty}^{+\infty},
\er
which again vanish at $x=\pm\infty$. Here, we have used explicitly the $t_3$ and $t_5$ time evolutions to show the conservation of the charges. 
However, it was already shown in \cite{Aratyn} that for every isospectral flow $t_N$ of the supersymmetric hierarchy, the charges derived from the spatial part of the
Lax operator $I_k^{(-n)}
$ are conserved. This is a novel property of the integrable hierarchy, and we will use it to consider the modified conserved quantities after introducing a defect in the theory.

\newpage
\subsection{Introducing defects}

Up to now, we have considered only conservation laws in the bulk theories. Now we will deal with the modification of the conserved quantities after introducing a 
defect placed at one particular point, say $x=0$. Let us start by considering the modification of the canonical  momentum for smKdV, which can be written as follows
\begin{equation}\label{P}
P=\int_{-\infty}^{0}dx\left[(\pt_x\phi_1)^2-i\bp_1\,\pt_x\bp_1\right]+\int_{0}^{+\infty}dx\left[(\pt_x\phi_2)^2-i\bp_2\,\pt_x\bp_2\right].
\end{equation}
Since we have already obtained the surface term in (\ref{dtP}), we have now non-zero contributions from the fields $\phi_1$, $\phi_2$ at the defect point, namely
\br\label{dtP1}
\frac{dP}{dt_3}&=&\frac{1}{4}\Big[2\pt_x\phi_1\pt^3_x\phi_1-3(\pt_x\phi_1)^4-(\pt^2_x\phi_1)^2+{9i}(\pt_x\phi_1)^2\bp_1\pt_x\bp_1+2i\pt_x\bp_1\pt_x^2\bp_1-i\bp_1\pt_x^3\bp_1\nonumber\\
&&-2\pt_x\phi_2\pt^3_x\phi_2+ 3(\pt_x\phi_2)^4+(\pt^2_x\phi_2)^2-9i(\pt_x\phi_2)^2\bp_2\pt_x\bp_2-2i\pt_x\bp_2\pt_x^2\bp_2+i\bp_2\pt_x^3\bp_2 \Big]_{x=0}.\nonu\\
\er
From eqs. (\ref{movu}) and (\ref{movpsi}), we also have that
\br\label{d3x}
\pt^3_x\phi_i&=&4\pt_{t_3}\phi_i+2(\pt_x\phi_i)^3-3i\, \pt_x\phi_i\, \bp_i\pt_x\bp_i,\quad\nonumber\\
\pt^3_x\bp_i&=&4\pt_{t_3}\bp_i+3\pt_x\phi_i\,\pt_x(\pt_x\phi_i\,\bp_i), \qquad \quad i=1,2,
\er
and  then eq. (\ref{dtP1}) becomes 
\br
\frac{dP}{dt_3} &=&\left[\, 2\pt_x\phi_1\,\pt_{t_3}\phi_1+\frac{1}{4}(\pt_x\phi_1)^4-\frac{1}{4}(\pt^2_x\phi_1)^2+i\bp_1\pt_{t_3}\bp_1+\frac{i}{2}\pt_x\bp_1\pt_x^2\bp_1\right.\nonumber\\
&&\left. \,\,- 2\pt_x\phi_2\,\pt_{t_3}\phi_2-\frac{1}{4}(\pt_x\phi_2)^4+\frac{1}{4}(\pt^2_x\phi_2)^2-i\bp_2\pt_{t_3}\bp_2-\frac{i}{2}\pt_x\bp_2\pt_x^2\bp_2\, \right]_{x=0}. \label{eq4.44}
\er
We point out that the canonical momentum is no longer conserved, since we have  field contributions at the defect point $x=0$. 
However, we still can use the defect conditions  (\ref{dxphi-})--(\ref{dxf1}), (\ref{dtphim}) and (\ref{dtf1}), to show that this contribution is a total $t_3$-derivative.
To do that, we can rewrite eq. (\ref{eq4.44}) in terms of the variables $\phi_{\pm}=\phi_1\pm\phi_2$ and $\bp_{\pm}=\bp_1\pm\bp_2$, as follows
\br
\frac{dP}{dt_3}&=&\left[\, \pt_x\phi_-\,\pt_{t_3}\phi_++\pt_x\phi_+\,\pt_{t_3}\phi_--\frac{1}{4}\pt^2_x\phi_-\,\pt^2_x\phi_++\frac{1}{8}(\pt_x\phi_-)^3\pt_x\phi_++\frac{1}{8}(\pt_x\phi_+)^3\pt_x\phi_-\right.\nonumber\\
&&\left.-\frac{i}{2}\bp_-\pt_{t_3}\bp_+-\frac{i}{2}\bp_+\pt_{t_3}\bp_-+\frac{i}{4}\pt_x\bp_-\pt^2_{x}\bp_++\frac{i}{4}\pt_x\bp_+\pt_x^2\bp_-\, \right]_{x=0}, \nonumber\\
&=&\left[\pt_{t_3}\left(\frac{4}{\o^2}\cosh\phi_+-\frac{2i}{\o}\cosh\left(\frac{\phi_+}{2}\right)f_1\bp_+\right)+\frac{2i}{\o}\cosh\left(\frac{\phi_+}{2}\right)\pt_{t_3}f_1\bp_+\right.\nonumber\\
&&\left. +\frac{i}{2\o^2}\left(\cosh^2\left(\frac{\phi_+}{2}\right)\bp_+\pt^2_x\bp_+-\frac{1}{4}\sinh(\phi_+)\pt_x\phi_+\bp_+\pt_x\bp_+\right)\right.\nonumber\\
&&\left.+\frac{6i}{\o^5}\sinh(\phi_+)\cosh^3\left(\frac{\phi_+}{2}\right)\pt_x\phi_+f_1\bp_+\,\right]_{x=0}.
\er
Then, by using eq. (\ref{dtf1}), we find that
\begin{equation}\label{Pdef}
\mathcal{P}=P-\Big[\frac{4}{\o^2}\cosh(\phi_+)-\frac{2i}{\o}\cosh\left(\frac{\phi_+}{2}\right)f_1\bp_+\Big]_{x=0},
\end{equation}
is the modified conserved momentum, which includes  defect contributions in order to preserve the original integrability, i.e. $  \frac{d\mathcal{P} }{dt_{3}}=0.   $\\

Now, the canonical energy in the presence of the defect is given by
\br\label{E}
E&=&\frac{1}{4}\int_{-\infty}^{0}dx\left[(\pt^2_x\phi_1)^2-(\pt_x\phi_1)^4+2\pt_x\phi_1\,\pt^3_x\phi_1-i\bp_1\,\pt_x^3\bp_1+3i(\pt_x\phi_1)^2\bp_1\pt_x\bp_1\right]\nonu\\
&&+\frac{1}{4}\int_{0}^{+\infty}dx\left[(\pt^2_x\phi_2)^2-(\pt_x\phi_2)^4+2\pt_x\phi_2\,\pt^3_x\phi_2-i\bp_2\,\pt_x^3\bp_2+3i(\pt_x\phi_2)^2\bp_2\pt_x\bp_2\right].\qquad\mbox{}
\end{eqnarray}
Considering only contributions at the defect as before, we get
\begin{eqnarray}
\frac{dE}{dt_3}&=&\left[E_1-E_2\right]_{x=0}
\end{eqnarray}
where
\br
E_{i} &=&\frac{1}{8}\pt_x\phi_i\,\pt^5_x\phi_i+\frac{1}{4}(\pt_x\phi_i)^6+\frac{1}{16}(\pt^3_x\phi_i)^2-\frac{3}{2}(\pt_x\phi_i)^2(\pt^2_x\phi_i)^2-(\pt_x\phi_i)^3\pt^3_x\phi_i\nonumber\\
&& -\frac{21i}{16}(\pt_x\phi_i)^4\bp_i\pt_x\bp_i +
\frac{15i}{16}(\pt^2_x\phi_i)^2\bp_i\pt_x\bp_i+\frac{27i}{16}\pt_x\phi_i\pt^3_x\phi_i\bp_i\pt_x\bp_i+\frac{3i}{2}\pt_x\phi_i\pt^2_x\phi_i\bp_i\pt^2_x\bp_i \nonumber\\
&& +\frac{3i}{4}(\pt_x\phi_i)^2\bp_i\pt^3_x\bp_i-\frac{i}{16}\bp_i\pt^5_x\bp_i+\frac{i}{16}\pt_x\bp_i\pt^4_x\bp_i-\frac{i}{16}\pt^2_x\bp_i\pt^3_x\bp_i,\qquad i=1,2.
\er
In order to obtain the defect contribution for the energy we take its $t_3$-derivative and use the equations of motion to get
\br
\frac{dE}{dt_3}\!\!&=&\!\!\left[\pt_{t_3}\phi_-\pt_{t_3}\phi_++\frac{1}{4}\pt_{t_3}\pt^2_x\phi_-\pt_x\phi_++\frac{1}{4}\pt_{t_3}\pt^2_x\phi_+\pt_x\phi_-+\frac{i}{8}\pt_x\bp_-\pt_{t_3}\pt_x\bp_++\frac{i}{8}\pt_x\bp_+\pt_{t_3}\pt_x\bp_-
\right.\nonumber\\&&+\left.
\frac{3i}{32}\left(\bp_-\pt_{t_3}\bp_++\bp_+\pt_{t_3}\bp_-\right)\left((\pt_x\phi_-)^2+(\pt_x\phi_+)^2\right)-\frac{i}{8}\pt^2_x\bp_-\pt_{t_3}\bp_+-\frac{i}{8}\pt^2_x\bp_+\pt_{t_3}\bp_-
\right.\nonumber\\
&&\left. +\frac{3i}{16}\pt_x\phi_-\pt_x\phi_+\left(\bp_-\pt_{t_3}\bp_-+\bp_+\pt_{t_3}\bp_+\right)-\frac{i}{8}\bp_-\pt_{t_3}\pt^2_x\bp_+-\frac{i}{8}\bp_+\pt_{t_3}\pt^2_x\bp_-\right]_{x=0}.
\er
Now, by using the B\"acklund equations (\ref{dxphi-})--(\ref{dxf1}), (\ref{dtphim}) and (\ref{dtf1}) we obtain
\br
\frac{dE}{dt_3}&=&\left[\pt_{t_3}\left(\frac{1}{\o^2}\pt^2_x\phi_+\sinh\phi_++\frac{1}{2\o^2}(\pt_x\phi_+)^2\cosh\phi_++\frac{6i}{\o^5}\cosh\Big(\frac{\phi_+}{2}\Big)\sinh^2\phi_+\,f_1\bp_+\right.\right.\nonu\\&&-\left.\left.\frac{i}{4\o}\sinh\Big(\frac{\phi_+}{2}\Big)\pt_x\phi_+f_1\pt_x\bp_+-\frac{i}{4\o}\sinh\Big(\frac{\phi_+}{2}\Big)\pt^2_x\phi_+f_1\bp_+-\frac{i}{2\o}\cosh\Big(\frac{\phi_+}{2}\Big)f_1\pt^2_x\bp_+\right)\right.\nonumber\\&&+\left.
\frac{i}{2\o}\sinh\Big(\frac{\phi_+}{2}\Big)\pt^2_x\phi_+\pt_{t_3}f_1\bp_+-\frac{i}{2\o}\sinh\Big(\frac{\phi_+}{2}\Big)\pt_x\phi_+\pt_{t_3}f_1\pt_x\bp_+ \right.\nonumber\\&&-\left.\frac{i}{2\o}\cosh\Big(\frac{\phi_+}{2}\Big)(\pt_x\phi_+)^2\pt_{t_3}f_1\bp_+-\frac{8}{\o^6}\sinh^3\phi_+\pt_{t_3}\phi_+ +\frac{i}{\o}\cosh\Big(\frac{\phi_+}{2}\Big)\pt_{t_3}f_1\pt^2_x\bp_+
\right.\nonumber\\&&+\left.\frac{12i}{\o^4}\cosh^2\Big(\frac{\phi_+}{2}\Big)\sinh\phi_+\pt_x\phi_+f_1\pt_{t_3}f_1-\frac{12i}{\o^5}\cosh\Big(\frac{\phi_+}{2}\Big)\sinh^2\phi_+\pt_{t_3}f_1\bp_+\right]_{x=0}\!\!.
\er
Finally using (\ref{dtf1}), we find that the modified conserved energy is given by
\br
\mathcal{E}&=&E-\Big[\frac{1}{\o^2}\left(\pt^2_x\phi_+\sinh\phi_++\frac{1}{2}(\pt_x\phi_+)^2\cosh\phi_+\right)+\frac{1}{\o^6}\left(6\cosh\phi_+-\frac{2}{3}\cosh(3\phi_+)\right)\nonu\\
&&\qquad \,-\frac{i}{2\o}\cosh\Big(\frac{\phi_+}{2}\Big)f_1\pt^2_x\bp_+-\frac{i}{4\o}\sinh\Big(\frac{\phi_+}{2}\Big)\left(\pt_x\phi_+f_1\pt_x\bp_++
\pt^2_x\phi_+f_1\bp_+\right)\nonu\\
&&\qquad \, +\frac{6i}{\o^5}\cosh\Big(\frac{\phi_+}{2}\Big)\sinh^2(\phi_+)f_1\bp_+\Big]_{x=0},
\er
i.e., $ \frac{d\mathcal{E} }{dt_{3}}=0$.
Although we have calculated the defect contributions to the modified conserved charges by considering conservation under $t_3$-derivative, it is natural to expect that this 
defect contributions are the same for all members of the integrable hierarchy and then, they are conserved under any $t_N$-derivative. In the appendix \ref{appF},
we check this statement explicitly for the $t_5$-derivative.

In general, to compute higher order modified conserved charges we can use the defect matrix in order to derive
a generating function of the defect contributions. To do that, let us consider the $K$ matrix (\ref{Kmatrix}), linking two different solutions of the linear problem (\ref{problemx}), (\ref{problemt}) by
\begin{eqnarray}\label{newsol}
\Omega^{(2)}=K\Omega^{(1)}.
\end{eqnarray}
Then, considering a defect located at $x=0$ we have for the first set of conserved quantities, 
\begin{align}\label{modified}
\mathcal{I}_1=\int_{-\infty}^{0}dx\left[V_{11}^{(1)}+V_{12}^{(1)}\G_{21}^{(1)}+V_{13}^{(1)}\G_{31}^{(1)}\right]
+\int_{0}^{+\infty}dx\left[V_{11}^{(2)}+V_{12}^{(2)}\G_{21}^{(2)}+V_{13}^{(2)}\G_{31}^{(2)}\right],
\end{align}
where $V_{ij}^{(p)}$ with $p=1,2$ are the spatial part of the Lax for each auxiliary problem (\ref{problemx}) in the region $x<0$ and $x>0$, and $\G_{21}^{(p)}=\Omega_{2}^{(p)}(\Omega_{1}^{(p)})^{-1}$ and $\G_{31}^{(p)}=\epsilon\,\Omega_{3}^{(p)}(\Omega_{1}^{(p)})^{-1}$ are the respective auxiliary functions for each region. Taking the time $t_N$-derivative of (\ref{modified}) and using the conservation equation (\ref{conservation}), we get
\begin{align}\label{dtmod}
\frac{d\mathcal{I}_1}{dt_N}=\left[U_{11}^{(1)}+U_{12}^{(1)}\G_{21}^{(1)}+U_{13}^{(1)}\G_{31}^{(1)}\right]_{x=0}-\left[U_{11}^{(2)}+U_{12}^{(2)}\G_{21}^{(2)}+U_{13}^{(2)}\G_{31}^{(2)}\right]_{x=0}
\end{align}
It is easy to see from (\ref{newsol}) that the auxiliary fields $\G_{21}^{(2)}$ and $\G_{31}^{(2)}$ satisfy the following relations,
\begin{align}
\G_{21}^{(2)}=\frac{K_{21}+K_{22}\G_{21}^{(1)}+K_{23}\G_{31}^{(1)}}{K_{11}+K_{12}\G_{21}^{(1)}+K_{13}\G_{31}^{(1)}},\qquad
\G_{31}^{(2)}=\frac{K_{31}+K_{32}\G_{21}^{(1)}+K_{33}\G_{31}^{(1)}}{K_{11}+K_{12}\G_{21}^{(1)}+K_{13}\G_{31}^{(1)}}.
\end{align}
Besides that, we have the gauge condition
\begin{equation}\label{dtnK}
\pt_{t_N}K=KA_{t_N}^{(1)}-A_{t_N}^{(2)}K=-KU^{(1)}+U^{(2)}K,
\end{equation}
Now, substituting these relations in (\ref{dtmod}) and using (\ref{dtnK}), (\ref{eq4.7}), (\ref{eq4.8}), we get
\begin{align}
\frac{d\mathcal{I}_1}{dt_N}=\left[-\frac{\pt_{t_N}\left(K_{11}+K_{12}\G_{21}^{(1)}+K_{13}\G_{31}^{(1)}\right)}{K_{11}+K_{12}\G_{21}^{(1)}+K_{13}\G_{31}^{(1)}}\right]_{x=0},
\end{align}
that the first generating function for the modified conserved quantities is $\mathcal{I}_1-\cD_1$, where 
\begin{align}\label{defect1}
\cD_1=-\ln\Big[K_{11}+K_{12}\G_{21}^{(1)}+K_{13}\G_{31}^{(1)}\Big],
\end{align}
is the defect contribution and depends on the elements of the defect matrix $K$. Then using the corresponding coefficients of expansions of the $\G_{21}$ and $\G_{31}$, we find that the lower terms can be written as follows,
\br
\cD_1^{(-1/2)}\!&=&\!\frac{2i}{\o}\, e^{\frac{\f_1+\f_2}{2}}\bp_1 f_1 + \frac{2}{\o^2}\, e^{\f_1+\f_2},\\
\cD_1^{(-1)}\!&=&\! \frac{i}{\o}\, e^{\frac{\f_1+\f_2}{2}}f_1\left(\bp_1\, \pt_x\f_1+\pt_x\bp_1\right)-\frac{2}{\o^2}\, e^{\f_1+\f_2}\, \pt_x\f_1-\frac{4i}{\o^3}\, e^{\frac{3(\f_1+\f_2)}{2}}f_1\bp_1 \nonumber\\&&+\frac{2}{\o^4}\, e^{2(\f_1+\f_2)},\qquad \,\,\mbox{}\\
\cD_1^{(-3/2)}&=&\frac{i}{2\o}\, e^{\frac{\f_1+\f_2}{2}}(\bp_1\, \pt^2_x\f_1+\pt_x\bp_1\pt_x\f_1+\pt^2_x\bp_1)f_1 +\frac{1}{\o^2}\, e^{\f_1+\f_2}\left[\pt^2_x\f_1+(\pt_x\f_1)^2\right]\nonu\\
&&-\frac{2i}{\o^3}\, e^{\frac{3(\f_1+\f_2)}{2}}(3\bp_1\, \pt_x\f_1+\pt_x\bp_1)f_1-\frac{4}{\o^4}\, e^{2(\f_1+\f_2)}\, \pt_x\f_1 -\frac{8i}{\o^5}\, e^{\frac{5(\f_1+\f_2)}{2}}f_1\bp_1 \nonu\\
&&+\frac{8}{3\o^6}\, e^{3(\f_1+\f_2)}.
\er
Analogously, we have the defect contributions for the second set of modified conserved quantities,
\br
 \cD_2=-\ln\Big[K_{22}+K_{21}\G_{12}^{(1)}+K_{23}\G_{32}^{(1)}\Big],
\er
from which we find explicitly the following coefficients,
\br
\cD_2^{(-1/2)}&=&\frac{2i}{\o}\, e^{\frac{-(\f_1+\f_2)}{2}}\bp_1f_1+\frac{2}{\o^2}\, e^{-(\f_1+\f_2)},\\
\cD_2^{(-1)}&=&\frac{i}{\o}\, e^{\frac{\f_1+\f_2}{2}}(\bp_1\, \pt_x\f_1-\pt_x\bp_1)f_1-\frac{2}{\o^2}\, e^{-(\f_1+\f_2)}\, \pt_x\f_1 +\frac{4i}{\o^3}\, e^{\frac{-3(\f_1+\f_2)}{2}}\bp_1 f_1\nonu\\&&+\frac{2}{\o^4}\, e^{-2(\f_1+\f_2)},\\
\cD_2^{(-3/2)}&=& \frac{i}{2\o}\, e^{\frac{-(\f_1+\f_2)}{2}}f_1(\bp_1\, \pt^2_x\f_1+\pt_x\bp_1\pt_x\f_1-\pt^2_x\bp_1)-\frac{1}{\o^2}\, e^{-(\f_1+\f_2)}\left[\pt^2_x\f_1-(\pt_x\f_1)^2\right]\nonu\\
&&+\frac{2i}{\o^3}\, e^{\frac{-3(\f_1+\f_2)}{2}}(3\bp_1\, \pt_x\f_1-\pt_x\bp_1)f_1+\frac{4}{\o^4}\, e^{-2(\f_1+\f_2)}\, \pt_x\f_1 +\frac{8i}{\o^5}\, e^{\frac{-5(\f_1+\f_2)}{2}}\bp_1f_1\nonu\\
&& + \frac{8}{3\o^6}\, e^{-3(\f_1+\f_2)}.
\er
We note that the defect contributions for the momentum
\begin{equation}
P_D=\cD_1^{(-1/2)}+\cD_2^{(-1/2)}=\Big[\frac{4}{\o^2}\cosh\phi_+-\frac{2 i}{\o}\cosh\left(\frac{\phi_+}{2}\right)f_1\bp_+\Big]_{x=0},
\end{equation} 
and for the energy
\br
E_D&=&\cD_1^{(-3/2)}+\cD_2^{(-3/2)}\nonu\\&=&\Big[\frac{1}{\o^2}\left(\pt^2_x\phi_+\sinh\phi_++\frac{1}{2}(\pt_x\phi_+)^2\cosh\phi_+\right)+\frac{1}{\o^6}\left(6\cosh\phi_+-\frac{2}{3}\cosh(3\phi_+)\right)\nonu\\
&&-\frac{i}{2\o}\cosh\left(\frac{\phi_+}{2}\right)f_1\pt^2_x\bp_+-\frac{i}{4\o}\sinh\left(\frac{\phi_+}{2}\right)\left(\pt_x\phi_+f_1\pt_x\bp_++
\pt^2_x\phi_+f_1\bp_+\right)\nonu\\
&&+\frac{6i}{\o^5}\cosh\left(\frac{\phi_+}{2}\right)\sinh^2(\phi_+)f_1\bp_+\Big]_{x=0},\qquad\,\, \mbox{}
\er 
can be properly recovered as linear combinations of the defect contributions $\cD_k^{(-n)}$, with $n=1/2$ and $n=3/2$ respectively. For sake of compactness, we have only considered contributions for
the canonical energy and momentum, however this method can be applied systematically to obtain higher  modified conserved charges of the ${\widehat sl}(2,1)$ integrable hierarchy. 
Furthermore, the generalization to other supersymmetric integrable hierarchies seems to be straightforward, and certainly deserve further investigations.


\section{Final remarks}
\label{sec5}

In this paper, we have studied the  presence of a type I integrable defects in the ${\widehat sl}(2,1)$ supersymmetric integrable hierarchy through super B\"acklund
transformations.
The invariance  of  the zero curvature representation  under gauge transformation allowed  the construction of a  defect-gauge matrix  connecting two different field
configurations of the same integrable model  and  hence generating the B\"acklund transformation.
The virtue of the method is that  all models within the hierarchy
are  constructed from the zero curvature representation and consequently are  all invariant under the  defect-gauge transformation.
It  therefore  leads to  a systematic  construction of  B\"acklund transformation  for all  models within the hierarchy in a universal manner.
Moreover,  in order to ensure the integrability, 
the presence of the defect requires  additional contributions to the  conserved quantities which were
constructed  from the defect   matrix in a systematic way.  We should point out that our arguments are general  and  the modified charges are conserved with respect to all flows.
To illustrate, we  have explicitly worked out  examples of B\"acklund transformation and  checked the conservation
for the first few charges  under flows
$t_3$ and $t_5$.

An interesting future application of  this framework  would be to generalize the construction to the super KdV hierarchy  as  proposed in \cite{Ana15}
for the pure bosonic case where  a Miura transformation  was realized in terms of gauge transformation.

Let us now remark that in our setting the integrable defect has been interpreted as frozen (space-like) B\"acklund transformations, which means that the field on the right of the defect can be generated from the field on the left of it by a B\"acklund transformation located at some fixed point, chosen here to be $x=0$. This is basically the definition of what is called type-I defect in the literature, which are associated to one soliton solution of the model. However, let us call the attention to an important property appearing in the corresponding supersymmetric extensions. It turns to be that what we are calling type-I integrable defect for the supersymmetric mKdV hierarchy contains intrinsically an auxiliary fermionic field necessary to describe defect conditions for the fermionic fields. In that sense, this kind of defect should be treated as a ``partial'' type-II defect, i.e. there is only one auxiliary fermionic time dependent
quantity defined on the defect, but not a bosonic auxiliary field. A genuine type-II defect will then contain one bosonic and two fermionic auxiliary fields as it is the case of the super sinh-Gordon model. Then, by using the universality argument,  the type-II super B\"acklund transformation for the smKdV can be obtained either directly from the type-II defect matrix for the super sinh-Gordon previously obtained in \cite{Nathaly}, or by applying the fusing procedure of two partial type I defect matrices. The latter procedure can be achieved  by performing two type-I B\"acklund transformations frozen at different points and then taking the limit when both points coincide. The auxiliary fields will then appear after an appropriate reparametrization of the ``squeezed'' fields valued only at the defect point. We expect that solutions for the auxiliary fields will be the same as those for the super sinh Gordon equation due to the universality of the spatial component of the Lax within the hierarchy, as it is in the bosonic case \cite{Ana_proceedings, Ana15}. We expect to return to these issues in future investigations.

On the other hand, there are also several issues concerning the classical soliton-defect interaction picture to be understood from both physical and mathematical point of view. First of all, the interaction of (multi)soliton solutions with defects can be determined by applying (several) B\"acklund transformations frozen at a fixed point, $x=0$. Now, it is well-known that this composition of transformations would produce in general multi-solitons solutions for an arbitrary non-linear integrable PDE (see \cite{Rogers} for a review). Then, a natural question one can ask is if two bulk integrable models (not necessarily the same) are defined on the left and right  of the defect, what is the physical effect after interaction between the defect and the soliton solutions?. Regarding to this issue, it has been shown that the preserving integrability condition requires that a single soliton solution passing through a defect suffers at most a delay, which depends on the defect (B\"acklund) parameter \cite{Corr1, Corr09, Parini}. In this case, the defect acts just as a purely transmitting interface, and  the physical effect of this inter\-action can be described for instance by studying the asymptotic behaviour of the auxiliary function by using the classical inverse scattering method (CISM) \cite{Faddeev}. This purely transmitting picture is also consistent with the quantum description \cite{Del1, Del2, Konik, Bowcock05, Zambon07} , where the interaction is described through infinite-dimensional transmission matrices determined by solving the quadratic triangle relations compatible with the bulk S-matrix, or the linear intertwining relations based on the infinite-dimensional representation of the quantum algebra associated to the model \cite{Corrigan:2010ph}.

Now, since the delay depends on the defect parameter value, it is also possible to have absorption or creation of solitons by the defect (see for instance \cite{Bowcock05, Parini, FLZ1}). In that case, besides interchanging, the defect can indeed store energy and momentum. And then, this feature suggests that if these integrable defects are implemented in physical systems then it would be probably an upper bound energy, a natural limit for the amount of energy that the defect will be able to store. Therefore, beyond that physical limit the system could radiate energy and then the integrability would be lost. Another important issue is that in such physical system the defect possessing  a non-zero momentum could in principle allows a time evolution of the defect degrees of freedom, even if it happens just around a small vicinity of the initial fixed point. The study of such effects can provide more information on how physical integrable defect interacts with soliton solutions. This kind of moving defects have been already discussed (see for instance \cite{Bowcock05}),  as well as some features related to the classical and quantum interaction picture have been addressed recently (see for instance \cite{Doikou2017}), however the complete physical and mathematical descriptions have not been totally established yet and remains as an interesting open problem to be considered in near future works.


\appendix
\section{Representation of the $ \widehat{\textit{sl}} $(2,1) affine Lie superalgebra}
\label{appA}

In this paper we are considering the following representation of the $\widehat{\textit{sl}} $(2,1) affine superalgebra,
\begin{equation}
K_1^{(2n+1)}=\begin{pmatrix}
0 & -\lambda^n & 0\\
-\lambda^{n+1} & 0 & 0\\
0 & 0 & 0
\end{pmatrix}, \quad K_2^{(2n+1)}=\begin{pmatrix}
\lambda^{n+\frac{1}{2}} & 0 & 0\\
0 & \lambda^{n+\frac{1}{2}} & 0\\
0 & 0 & 2\lambda^{n+\frac{1}{2}}
\end{pmatrix},
\end{equation}
\begin{equation}
M_1^{(2n+1)}=\begin{pmatrix}
0 & -\lambda^n & 0\\
\lambda^{n+1} & 0 & 0\\
0 & 0 & 0
\end{pmatrix},\quad
M_2^{(2n)}=\begin{pmatrix}
\lambda^n & 0 & 0\\
0 & -\lambda^n & 0\\
0 & 0 & 0
\end{pmatrix},
\end{equation}
\begin{equation}
F_1^{\left(2n+\frac{3}{2}\right)}=\begin{pmatrix}
0 & 0 & \lambda^{n+\frac{1}{2}}\\
0 & 0 & -\lambda^{n+1} \\
\lambda^{n+1} & -\lambda^{n+\frac{1}{2}} & 0
\end{pmatrix},\quad F_2^{\left(2n+\frac{1}{2}\right)}=\begin{pmatrix}
0 & 0 & -\lambda^{n}\\
0 & 0 & \lambda^{n+\frac{1}{2}} \\
\lambda^{n+\frac{1}{2}} & -\lambda^{n} & 0
\end{pmatrix},
\end{equation}
\begin{equation}
G_1^{\left(2n+\frac{1}{2}\right)}=\begin{pmatrix}
0 & 0 & \lambda^{n}\\
0 & 0 & \lambda^{n+\frac{1}{2}} \\
\lambda^{n+\frac{1}{2}} & \lambda^{n} & 0
\end{pmatrix},\quad G_2^{\left(2n+\frac{3}{2}\right)}=\begin{pmatrix}
0 & 0 & -\lambda^{n+\frac{1}{2}}\\
0 & 0 & -\lambda^{n+1} \\
\lambda^{n+1} & \lambda^{n+\frac{1}{2}} & 0
\end{pmatrix}.
\end{equation}


\section{$N=5$ Lax component}
\label{appB}
The Lax component $A_{t_5}$ takes the following form,%

\begin{equation}
\renewcommand{\arraystretch}{1.5}
A_{t_5}=\left(
\begin{array}{cc|c}
b_{11} \,& b_{12} \,& \,\,b_{13}\\
b_{21} \,& b_{22} \,& \,\,b_{23}\\\hline
b_{31} \,& b_{32} \,& \,\,b_{33}
\end{array}
\right),
\label{At5}
\end{equation}
where,
\br
 b_{11}&=&\l^{5/2}+\l^2\,\dxx-\frac{i\l^{3/2}}{2}\bp\pt_x\bp+\l\left(\frac{1}{2}(\dx)^3-\frac{1}{4}\dxxx-\frac{3i}{4}\dx\,\bp\pt_x\bp\right)\nonu\\
&& +\l^{1/2}\left(\frac{i}{2}(\dx)^2\,\bp\pt_x\bp+\frac{i}{8}\pt_x\bp\pt^2_x\bp-\frac{i}{8}\bp\pt^3_x\bp \right)+\frac{5}{8}(\dx)^2\dxxx+\frac{5}{8}\dx(\dxx)^2\nonu\\
&& +\frac{5i}{4}(\dx)^3\,\bp\pt_x\bp-\frac{5i}{16}\dx\,\bp\pt_x^3\bp-\frac{5i}{16}\dxx\,\bp\pt_x^2\bp-\frac{5i}{16}\dxxx\,\bp\pt_x\bp -\frac{3}{8}(\dx)^5,\\[8pt]       
 b_{12}&=&-\l^2+\frac{\l}{2}\left(\dxx+(\dx)^2-i\bp\pt_x\bp\right)-\frac{1}{8}(\dxx)^2+\frac{1}{8}\dxxxx+\frac{1}{4}\dx\dxxx-\frac{3}{4}(\dx)^2\dxx\nonu\\
&&-\frac{3}{8}(\dx)^4+\frac{i}{4}\dx\bp\pt_x^2\bp+i(\dx)^2\bp\pt_x\bp+\frac{i}{2}\dxx\,\bp\pt_x\bp+\frac{i}{8}\pt_x\bp\pt_x^2\bp-\frac{i}{8}\bp\pt_x^3\bp,\qquad\mbox{}\\[8pt] 
 b_{13}&=&\l^2\sqrt{i}\bp+\frac{\l^{3/2}\sqrt{i}}{2}\left(\dx\,\bp+\pt_x\bp\right)+\frac{\l\sqrt{i}}{4}\left(\dx\pt_x\bp- 2(\dx)^2\bp-\dxx\bp + \pt_x^2\bp\right)\nonu\\
&&+\frac{\l^{1/2}\sqrt{i}}{8}\left(\dx\pt_x^2\bp-3(\dx)^2\pt_x\bp-\dxx\pt_x\bp+\dxxx\bp-3\dx\dxx\bp-3(\dx)^3\bp+\pt_x^3\bp\right)\nonu\\
&&+\frac{\sqrt{i}}{16}\left(\dx\,\pt_x^3\bp-\dxx\,\pt_x^2\bp+\dxxx\,\pt_x\bp-\dxxxx\,\bp+\pt_x^4\bp\right)-\frac{\sqrt{i}}{2}\dx\dxx\pt_x\bp\nonu\\
&&+\frac{\sqrt{i}}{4}\left(\dxx\,(\dx)^2\,\bp-(\dx)^3\pt_x\bp-(\dx)^2\,\pt_x^2\bp\right) \nonumber \\&&+\frac{\sqrt{i}}{8}\left(3(\dx)^4\,\bp-3\dx\,\dxxx\,\bp-(\dxx)^2\bp\right),\\[8pt]
 b_{21}&=&-\l^3+\frac{\l^2}{2}\left(-\dxx+(\dx)^2-i\bp\pt_x\bp\right)+\l\left(\frac{1}{4}\dx\,\dxxx-\frac{1}{8}(\dxx)^2-\frac{1}{8}\dxxxx\right.\nonu\\
&&\left. -\frac{3}{8}(\dx)^4+\frac{3}{4}(\dx)^2\dxx+i(\dx)^2\bp\pt_x\bp-\frac{i}{4}\dx\bp\pt_x^2\bp-\frac{i}{2}\dxx\bp\pt_x\bp  \nonumber \right. \\&&\left.  +\frac{i}{8}\pt_x\bp\pt_x^2\bp-\frac{i}{8}\bp\pt_x^3\bp\right),\\[8pt]
b_{22}&=&\l^{5/2}+\l^2\dx-\frac{i\l^{3/2}}{2}\bp\pt_x\bp+\l\left(\frac{1}{4}\dxxx-\frac{1}{2}(\dx)^3+\frac{3i}{4}\dx\bp\pt_x\bp\right)\nonu\\
&&+\l^{1/2}\left(\frac{i}{2}(\dx)^2\bp\pt_x\bp+\frac{i}{8}\pt_x\bp\pt_x^2\bp-\frac{i}{8}\bp\pt_x^3\bp\right)+\frac{1}{16}\dxxxxx-\frac{5}{8}\dx(\dxx)^2\nonu\\
&& -\frac{5}{8}\dxxx(\dx)^2+\frac{3}{8}(\dx)^5+\frac{5i}{16}\dxx\bp\pt_x^2\bp+\frac{5i}{16}\dxxx\bp\pt_x\bp+\frac{5i}{16}\dx\bp\pt_x^3\bp \nonumber\\&&-\frac{5i}{4}(\dx)^3\bp\pt_x\bp,
\er
\br
b_{23}&=&\l^{5/2}\sqrt{i}\bp+\frac{\l^2\sqrt{i}}{2}\left(\pt_x\bp-\dx\bp\right)+\frac{\l^{3/2}\sqrt{i}}{4}\left(\dxx\bp-2(\dx)^2\bp-\dx\pt_x\bp+\pt_x^2\bp\right)\nonu\\
&&+\frac{\l\sqrt{i}}{8}\left(\dxx\pt_x\bp-\dx\pt_x^2\bp-\dxxx\bp-3(\dx)^2\pt_x\bp+3(\dx)^3\bp-3\dx\dxx\bp+\pt_x^3\bp\right)\nonu\\
&&+\frac{\l^{1/2}\sqrt{i}}{16}\left(\dxxxx\,\bp-\dxxx\,\pt_x\bp-\dx\,\pt_x^3\bp+\dxx\,\pt_x^2\bp+\pt_x^4\bp\right)-\frac{\l^{1/2}\sqrt{i}}{2}\dx\,\dxx\,\pt_x\bp\nonu\\
&&+\frac{\l^{1/2}\sqrt{i}}{4}\left((\dx)^3\pt_x\bp-(\dx)^2\pt_x^2\bp-\dxx(\dx)^2\bp+\frac{3}{2}(\dx)^4\,\bp- 12\dxxx\,\dx\,\bp \right. \nonu\\&&\left. \qquad \qquad\,\,\, - 4(\dxx)^2\,\bp\right),\\[8pt]
 b_{31}&=&\l^{5/2}\sqrt{i}\bp+\frac{\l^2\sqrt{i}}{2}(\dx\bp-\pt_x\bp)+\frac{\l^{3/2}\sqrt{i}}{4}\left(\dxx\bp-\dx\pt_x\bp-2(\dx)^2\bp+\pt_x^2\bp\right)\nonu\\
&&+\frac{\l\sqrt{i}}{8}\left(\dx\pt_x^2\bp-\dxx\pt_x\bp+\dxxx\bp+3\dx\dxx\bp-3(\dx)^3\bp+3(\dx)^2\pt_x\bp-\pt_x^3\bp\right)\nonu\\
&&+\frac{\l^{1/2}\sqrt{i}}{16}\left(\dxxxx\,\bp-\dxxx\,\pt_x\bp-\dx\pt_x^3\bp+\dxx\,\pt_x^2\bp+\pt_x^4\bp\right)-\frac{\l^{1/2}\sqrt{i}}{2}\dx\,\dxx\,\pt_x\bp\nonu\\
&&+\frac{\l^{1/2}\sqrt{i}}{4}\left((\dx)^3\pt_x\bp-(\dx)^2\pt_x^2\bp-\dxx(\dx)^2\bp-\frac{1}{2}(\dxx)^2\bp- 12\dx\dxxx\bp \right. \nonu \\&&\left. \qquad \qquad \,\,\,+12(\dx)^4\,\bp\right),
\er
\br
 b_{32}&=&\l^2\sqrt{i}\bp-\frac{\l^{3/2}\sqrt{i}}{2}(\dx\,\bp+\pt_x\bp)+\frac{\l\sqrt{i}}{4}\left(\dx\pt_x\bp-\dxx\,\bp-2(\dx)^2\,\bp+\pt_x^2\bp\right)\nonu\\
&&+\frac{\l^{1/2}\sqrt{i}}{8}\left(3\dx\dxx\bp+3(\dx)^3\bp-\dxxx\bp+\dxx\pt_x\bp-\dx\pt_x^2\bp+3(\dx)^2\pt_x\bp-\pt_x^3\bp\right)\nonu\\
&&+\frac{\sqrt{i}}{16}\left(\dx\,\pt_x^3\bp-\dxx\,\pt_x^2\bp-\dxxxx\,\bp+\dxxx\,\pt_x\bp+\pt_x^4\bp\right)-\frac{\sqrt{i}}{2}\dx\,\dxx\,\pt_x\bp\nonu\\
&&-\frac{\sqrt{i}}{4}(\dx)^2\,\pt_x^2\bp-\frac{\sqrt{i}}{4}(\dx)^3\,\pt_x\bp-\frac{\sqrt{i}}{8}(\dxx)^2\,\bp-\frac{3}{8}\dxxx\,\dx\,\bp+\frac{\sqrt{i}}{4}(\dx)^2\,\dxx\,\bp \nonumber \\&&+\frac{3}{8}(\dx)^4\,\bp,\\[8pt]
 b_{33}&=&2\l^{5/2}-i\l^{3/2}\bp\pt_x\bp+i\l^{1/2}\left((\dx)^2\,\bp\pt_x\bp+\frac{1}{4}\pt_x\bp\pt_x^2\bp-\frac{1}{4}\bp\pt_x^3\bp\right).      
\er


\section{Coefficients of the B\"acklund transformations for $N=5$ member}
\label{appC}

The coefficients $c_i$ in B\"acklund equations (\ref{btt5phi}) are given by,
\br
c_0&=&-\pt^4_x\phi_+\cosh\Big(\frac{\phi_+}{2}\Big)+\left(\pt^2_x\phi_+\right)^2\sinh\Big(\frac{\phi_+}{2}\Big)+3\left(\pt^3_x\phi_+\right)\left(\pt_x\phi_+\right)\sinh\Big(\frac{\phi_+}{2}\Big)\nonu\\&&+
\left(\pt^2_x\phi_+\right)\left(\pt_x\phi_+\right)^2\cosh\Big(\frac{\phi_+}{2}\Big)-\frac{3}{4}\left(\pt_x\phi_+\right)^4\sinh\Big(\frac{\phi_+}{2}\Big),\label{c0}\\
c_1&=&\pt^3_x\phi_+\cosh\Big(\frac{\phi_+}{2}\Big)-\left(\pt_x\phi_+\right)^3\cosh\Big(\frac{\phi_+}{2}\Big)+4\left(\pt^2_x\phi_+\right)\left(\pt_x\phi_+\right)\sinh\Big(\frac{\phi_+}{2}\Big),\\
c_2&=&-\left(\pt^2_x\phi_+\right)\cosh\Big(\frac{\phi_+}{2}\Big)+2\left(\pt_x\phi_+\right)^2\sinh\Big(\frac{\phi_+}{2}\Big),\\
c_3&=&\pt_x\phi_+\cosh\Big(\frac{\phi_+}{2}\Big),\\
c_4&=&-2\sinh\Big(\frac{\phi_+}{2}\Big),\\
c_5&=&4\left(\pt^4_x\phi_+\right)\cosh\phi_+-6\left(\pt^2_x\phi_+\right)\left(\pt_x\phi_+\right)^2\cosh\phi_++2\left(\pt^2_x\phi_+\right)^2\sinh\phi_+ \nonu\\&&-4\left(\pt^3_x\phi_+\right)\left(\pt_x\phi_+\right)\sinh\phi_++\frac{3}{2}\left(\pt_x\phi_+\right)^4\sinh\phi_+,\\
c_6&=& 4\left(\pt^2_x\phi_+\right)\cosh\phi_+-4\left(\pt_x\phi_+\right)^2\sinh\phi_+,\\
c_7&=&2(\pt_x\phi_+)(\cosh\phi_+),\\
c_8&=&2\sinh\phi_+,\qquad \mbox{}\\
c_9&=&\left[	-20\cosh\left(\frac{\phi_+}{2}\right)+20\cosh\left(\frac{3\phi_+}{2}\right)+80\cosh\left(\frac{5\phi_+}{2}\right)\right](\pt^2_x\phi_+)\nonumber\\
	&& +\left[\,35\,\sinh\left(\frac{\phi_+}{2}\right)-\frac{15}{2}\sinh\left(\frac{3	\phi_+}{2}\right)+\frac{75}{2}\sinh\left(\frac{5\phi_+}{2}\right)\right](\pt_x\phi_+)^2,\\
c_{10}&=&\left[70\cosh\left(\frac{\phi_+}{2}\right)-25\cosh\left(\frac{3\phi_+}{2}\right)+35\cosh\left(\frac{5\phi_+}{2}\right)\right]\pt_x\phi_+,\\
c_{11}&=&-20\sinh\left(\frac{\phi_+}{2}\right)+10\sinh\left(\frac{3\phi_+}{2}\right)+30\sinh\left(\frac{5\phi_+}{2}\right),\\
c_{12}&=&40\pt^2_x\phi_+\left(\cosh\phi_+-\cosh (3\phi_+)\right)-20(\pt_x\phi_+)^2\left(5\sinh\phi_++\sinh(3\phi_+)\right),\\
c_{13}&=&30\sinh\phi_+-10\sinh(3\phi_+),\\
c_{14}&=&-120\sinh\left(\frac{\phi_+}{2}\right)+80\sinh\left(\frac{3\phi_+}{2}\right)+240\sinh\left(\frac{5\phi_+}{2}\right)-60\sinh\left(\frac{7\phi_+}{2}\right)\nonu\\&&-100\sinh\left(\frac{9\phi_+}{2}\right),\\
c_{15}&=&240\sinh\phi_+-120\sinh(3\phi_+)+24\sinh(5\phi_+).\label{c15}
\er

\noindent And the coefficients $g_j$ in the B\"acklund equations (\ref{btt5psi}) are the following,
\br
g_0&=&\Big[-\frac{1}{2}(\pt^2_x\phi_+)^2-\frac{3}{2}(\pt^3_x\phi_+)(\pt_x\phi_+)+\frac{3}{8}(\pt_x\phi_+)^4\Big]\cosh\Big(\frac{\phi_+}{2}\Big)\nonumber\\
	&& +\Big[\frac{1}{2}\pt^4_x\phi_+-\frac{1}{2}(\pt^2_x\phi_+)(\pt_x\phi_+)^2\Big]\sinh\Big(\frac{\phi_+}{2}\Big),\label{d0}
\er
\br	
g_1&=&\left[-\frac{1}{2}\pt^3_x\phi_++\frac{1}{2}(\pt_x\phi_+)^3\right]\sinh\Big(\frac{\phi_+}{2}\Big)-2(\pt^2_x\phi_+)(\pt_x\phi_+)\cosh\Big(\frac{\phi_+}{2}\Big),\\
g_2&=&\frac{1}{2}(\pt^2_x\phi_+)\sinh\Big(\frac{\phi_+}{2}\Big)-(\pt_x\phi_+)^2\cosh\Big(\frac{\phi_+}{2}\Big),\\
g_3&=&-\frac{1}{2}\pt_x\phi_+\sinh\Big(\frac{\phi_+}{2}\Big),\\
g_4&=&\cosh\Big(\frac{\phi_+}{2}\Big),\\
g_5&=&\left[10\sinh\phi_+-5\sinh(2\phi_+)\right]\pt_x\phi_+,\\
g_6&=&-\left[\frac{5}{2}+20\cosh\phi_+	+\frac{35}{2}\cosh(2\phi_+)\right](\pt^2_x\phi_+)(\pt_x\phi_+)\nonu\\
&&-\left[10\sinh\phi_+ +5\sinh(2\phi_+)\right]\pt^3_x\phi_+ -\left[\frac{15}{2}\sinh\phi_+ +\frac{15}{4}\sinh(2\phi_+)\right](\pt_x\phi_+)^3, \qquad \mbox{}\\
g_7&=&-\left[\frac{35}{2}\cosh\left(\frac{\phi_+}{2}\right)+\frac{45}{4} \cosh\left(\frac{3\phi_+}{2}\right)+\frac{45}{4}\cosh\left(\frac{5\phi_+}{2}\right)\right](\pt_x\phi_+)^2\nonumber\\
	&&-\left[15\sinh\left(\frac{3\phi_+}{2}\right)+15\sinh\left(\frac{5\phi_+}{2}\right)\right]\pt^2_x\phi_+,\\
g_8&=&\left[-15\sinh\left(\frac{\phi_+}{2}\right)-\frac{45}{2}\sinh\left(\frac{3\phi_+}{2}\right)-\frac{15}{2}\sinh\left(\frac{5\phi_+}{2}\right)\right](\pt_x\phi_+),\\
g_9&=&10\cosh\left(\frac{\phi_+}{2}\right)-5\cosh\left(\frac{3\phi_+}{2}\right)-5\cosh\left(\frac{5\phi_+}{2}\right),\\
g_{10}&=&\left[-120\sinh\phi_+-40\sinh(2\phi_+)+40\sinh(3\phi_+)+20\sinh(4\phi_+)\right]\pt_x\phi_+,\\
g_{11}&=&60\cosh\left(\frac{\phi_+}{2}\right)-40\cosh\left(\frac{3\phi_+}{2}\right)-40\cosh\left(\frac{5\phi_+}{2}\right)+10\cosh\left(\frac{7\phi_+}{2}\right)\nonumber\\
	&& +10\cosh\left(\frac{9\phi_+}{2}\right)\label{d11}.
\er


\section{Derivation of the super-B\"acklund transformation for $N=3$ and $N=5$ super equations}
\label{appD}

We propose that the spatial part of the super-B\"acklund transformation for smKdV hierarchy takes the following form,
\br
  D\Phi_- &=& \frac{4i}{\omega}\cosh\Big(\frac{\Phi_+}{2}\Big)\S , \label{asb1}\\
  D\S &=& -\frac{2i}{\omega} \sinh\Big(\frac{\Phi_+}{2}\Big),\label{asb3}
\er
with $\S$  being the fermionic superfield introduced in section \ref{super}. Now, from eq (\ref{DPhi}), namely
\br 
 D_{t_3} \Phi = -D^6\Phi + 2(D^2\Phi)^3 - 3 (D\Phi)(D^2\Phi)(D^3\Phi),
\er
we get the following relation for the temporal ($t_3$)  part of the transformation, 
\br
 D_{t_3}\Phi_- =  T_1 +T_2 +T_3, \label{termos}
\er
where the terms $T_1$, $T_2$ and $T_3$ in eq. (\ref{termos}) are given as follows,
\br
T_1&=& -D^6\Phi_-\nonumber\\ &=& -\frac{i}{\omega}\cosh\Big(\frac{\Phi_+}{2}\Big)\left[D^4\Phi_+ D\Phi_+ + 2 D^2\Phi_+D^3\Phi_+\right]\S,\nonumber\\
&& -\frac{i}{2\omega}\sinh\Big(\frac{\Phi_+}{2}\Big)\left[(D^2\Phi_+)^2 D\Phi_+ + 4 D^5\Phi_+\] \S\nonumber \\
&&-\frac{1}{\omega^2} \sinh\Phi_+ \left[4(D^2\Phi_+)^2 + D^3\Phi_+ D\Phi_+ \]  -\frac{4}{\o^2}\cosh\Phi_+ (D^4\Phi_+),\\[0.3cm]
T_2&=&2\left[(D^2\Phi_1)^3-(D^2\Phi_2)^3\] \nonumber\\
&=& \frac{6}{\omega^2}\sinh\Phi_+ (D^2\Phi_+)^2 +\frac{3i}{\omega}\sinh\Big(\frac{\Phi_+}{2}\Big)(D^2\Phi_+)^2(D\Phi_+)\S \nonumber \\
&&+\frac{32}{\o^6}\sinh^3\Phi_+ +\frac{48i}{\o^5}\sinh^2\Phi_+\sinh\Big(\frac{\Phi_+}{2}\Big) (D\Phi_+)\S,
\\[0.3cm]
T_3 &=& -3\left[(D\Phi_1)(D^2\Phi_1)(D^3\Phi_1)-(D\Phi_2)(D^2\Phi_2)(D^3\Phi_2)\]\nonumber\\
&=& -\frac{3}{\o^2}\sinh\Phi_+(D\Phi_+)(D^3\Phi_+) +\frac{3i}{\o}\cosh\Big(\frac{\Phi_+}{2}\Big)(D^2\Phi_+)(D^3\Phi_+) \S \nonumber \\
&&- \frac{3i}{2\o}\sinh\Big(\frac{\Phi_+}{2}\Big) (D\Phi_+)(D^2\Phi_+)^2\S \nonumber \\
&&+\frac{48i}{\o^5}\sinh\Phi_+ \cosh\Phi_+ \cosh\Big(\frac{\Phi_+}{2}\Big) (D\Phi_+)\S \nonumber \\
&&- \frac{24i}{\o^5}\sinh^2\Phi_+\sinh\Big(\frac{\Phi_+}{2}\Big)(D\Phi_+)\S.
\er
Then, we find that
\br
 D_{t_3}\Phi_- &=& \frac{2}{\o^2} \sinh\Phi_+\left[(D^2\Phi_+)^2-(D\Phi_+)(D^3\Phi_+)\] -\frac{4}{\o^2}\cosh\Phi_+ (D^4\Phi_+) +\frac{32}{\o^6}\sinh^3\Phi_+\nonumber \\[0.1cm]
 &&+\frac{i}{\omega}\cosh\Big(\frac{\Phi_+}{2}\Big)\left[D^2\Phi_+D^3\Phi_+ -D^4\Phi_+ D\Phi_+\right]\S,\nonumber\\[0.1cm]
&&+\frac{i}{\o}\sinh\Big(\frac{\Phi_+}{2}\Big)\left[(D^2\Phi_+)^2(D\Phi_+) - 2 D^5\Phi_+\]\S \nonumber\\
&&+\frac{96i}{\o^5}\left[\sinh\Big(\frac{\Phi_+}{2}\Big)+4\sinh^3\Big(\frac{\Phi_+}{2}\Big)+3\sinh^5\Big(\frac{\Phi_+}{2}\Big)\](D\Phi_+)\S. \label{asb2}
\er
Now, by imposing consistency of eqs. (\ref{asb1}) and (\ref{asb2}), namely $D_{t_3}D\Phi_- = DD_{t_3} \Phi_-$, we found that
\br
 D_{t_3} \S &=& -\frac{i}{2\o}\cosh\Big(\frac{\Phi_+}{2}\Big)\left[(D\Phi_+)(D^2\Phi_+)^2-2(D^5\Phi_+)\]\nonumber\\
 &&-\frac{i}{2\o}\sinh\Big(\frac{\Phi_+}{2}\Big)\left[(D^2\Phi_+)(D^3\Phi_+)-(D\Phi_+)(D^4\Phi_+)\]\nonumber \\[0.1cm]
 &&+\frac{12}{\o^4} \sinh\Phi_+\cosh^2\Big(\frac{\Phi_+}{2}\Big)(D^2\Phi_+)\S -\frac{12i}{\o^5}\sinh^2\Phi_+\cosh\Big(\frac{\Phi_+}{2}\Big) (D\Phi_+) . \label{asb4}
\er
Finally, we have found that the super-B\"acklund transformation for the smKdV is given by equations (\ref{asb1}), (\ref{asb3}), (\ref{asb2}) and (\ref{asb4}).

Let us now construct the supersymmetric extension for the temporal part of the super B\"acklund transformation for the super $t_5$-equation. Using eq. (\ref{Dt5Phi}), we get
\br
 D_{t_5} \Phi_- &=&  S_1 +S_2+S_3+S_4+S_5+S_6+S_7+S_8,
\er
where
\br 
 S_1&=& D^{10} \Phi_-\nonumber \\
 &=& \frac{i}{\omega}\cosh\Big(\frac{\Phi_+}{2}\Big)\left[ \frac{3}{2}(D^2\Phi_+)^2(D^4\Phi_+)(D\Phi_+) +(D^2\Phi_+)^3(D^3\Phi_+)+4(D^2\Phi_+)(D^7\Phi_+)     \right. \nonumber \\  && \left. \qquad \qquad \qquad \,\,\,+6(D^4\Phi_+)(D^5\Phi_+) +4 (D^6\Phi_+)(D^3\Phi_+) + (D^8\Phi_+)(D\Phi_+)\right] \Sigma\nonumber\\&&+\frac{i}{\omega}\sinh\Big(\frac{\Phi_+}{2}\Big)\left[\frac{1}{8}(D^2\Phi_+)^4(D\Phi_+) +3 (D^2\Phi_+)^2(D^5\Phi_+) +2(D^2\Phi_+)(D^6\Phi_+)(D\Phi_+) \right. \nonumber \\&&\left.\qquad \qquad \qquad\quad +6 (D^2\Phi_+)(D^4\Phi_+)(D^3\Phi_+)+\frac{3}{2}(D^4\Phi_+)^2(D\Phi_+)+2(D^9\Phi_+)\right] \Sigma \nonumber\\
 &&+\frac{1}{\omega^2}\sinh\Phi_+\left[12(D^4\Phi_+)^2 +16(D^2\Phi_+)(D^6\Phi_+) +3 (D^7\Phi_+)(D\Phi_+) \nonumber \right. \\&& \qquad \qquad \qquad \left. +2(D^5\Phi_+)(D^3\Phi_+) + 4 (D^2\Phi_+)^4 +\frac{11}{4}(D^2\Phi_+)^2(D^3\Phi_+)(D\Phi_+)\right]\nonumber \\
 && +\frac{1}{\omega^2}\cosh\Phi_+ \left[24(D^2\Phi_+)^2(D^4\Phi_+)+\frac{11}{2}(D^2\Phi_+)(D^5\Phi_+)(D\Phi_+) +4(D^8\Phi_+) \right. \nonu \\ &&\qquad \qquad \qquad \left. +\frac{7}{2}(D^4\Phi_+)(D^3\Phi_+)(D\Phi_+)\right] \nonumber\\
 &&+\frac{5}{2\omega^2}\left[(D^4\Phi_+)(D^3\Phi_+)(D\Phi_+)+(D^2\Phi_+)(D^5\Phi_+)(D\Phi_+)\right],\\[1cm]
S_2&=& 5\left[(D\Phi_1)(D^2\Phi_1)(D^7\Phi_1)-(D\Phi_2)(D^2\Phi_2)(D^7\Phi_2) \right]\nonu\\
&=& \frac{5i}{4\omega}\cosh\Big(\frac{\Phi_+}{2}\Big)\left[3 (D^2\Phi_+)^2(D^4\Phi_+)(D\Phi_+)-4 (D^2\Phi_+)(D^7\Phi_+)\right] \Sigma \nonu\\
&&+\frac{5i}{8\omega}\sinh\Big(\frac{\Phi_+}{2}\Big)\left[(D^2\Phi_+)^4(D\Phi_+)+4(D^2\Phi_+)(D^6\Phi_+)(D\Phi_+)\right]\Sigma\nonu \\
&&+\frac{5}{4\omega^2} \sinh\Phi_+\left[5(D^2\Phi_+)^2(D\Phi_+)(D^3\Phi_+) +4 (D\Phi_+)(D^7\Phi_+)\right]\nonu\\&&+\frac{5}{\omega^2}\cosh^2\Big(\frac{\Phi_+}{2}\Big)(D\Phi_+)(D^2\Phi_+)(D^5\Phi_+)\nonu\\
&& -\frac{5i}{\omega^5}\cosh\Big(\frac{\Phi_+}{2}\Big)\sinh(2\Phi_+)\left[4(D^5\Phi_+)+7(D^2\Phi_+)^2(D\Phi_+)\right]\Sigma \nonu \\
&& -\frac{20i}{\omega^5}\cosh\Big(\frac{\Phi_+}{2}\Big)\sinh^2\Phi_+\left[5(D^2\Phi_+)(D^3\Phi_+)+4(D^4\Phi_+)(D\Phi_+)\right]\Sigma\nonu\\
&& -\frac{10i}{\omega^5}\cosh\Big(\frac{\Phi_+}{2}\Big)\sinh\Phi_+\left[(D^2\Phi_+)^2(D\Phi_+)+4(D^5\Phi_+)\right]\Sigma,
\er
\br 
S_3 &=& 5\left[(D\Phi_1)(D^3\Phi_1)(D^6\Phi_1)-(D\Phi_2)(D^3\Phi_2)(D^6\Phi_2) \right] \nonu\\
&=& \frac{5i}{2\omega}\sinh\Big(\frac{\Phi_+}{2}\Big)\left[(D\Phi_+)(D^3\Phi_+)(D^5\Phi_+)+(D\Phi_+)(D^2\Phi_+)(D^6\Phi_+)\right]\Sigma \nonu\\
&&-\frac{5i}{\omega}\cosh\Big(\frac{\Phi_+}{2}\Big) (D^3\Phi_+)(D^6\Phi_+)\Sigma\\
&&+\frac{5}{\omega^2}\sinh\Phi_+(D\Phi_+)(D^2\Phi_+)^2(D^3\Phi_+) + \frac{5}{\omega^2}\cosh\Phi_+(D\Phi_+)(D^3\Phi_+)(D^4\Phi_+)\nonu\\
&&-\frac{80i}{\omega^5}\cosh^3\Big(\frac{\Phi_+}{2}\Big)\sinh\Phi_+ (D\Phi_+)(D^2\Phi_+)^2\Sigma \nonumber\\&&-\frac{80i}{\omega^5}\cosh^3\Big(\frac{\Phi_+}{2}\Big)\cosh\Phi_+ (D\Phi_+)(D^4\Phi_+)\Sigma\nonu,\\[1cm]
S_4 &=& 5\left[(D\Phi_1)(D^4\Phi_1)(D^5\Phi_1)-(D\Phi_2)(D^4\Phi_2)(D^5\Phi_2) \right] \nonu\\
&=&
\frac{5i}{2\omega}\sinh\Big(\frac{\Phi_+}{2}\Big)\left[(D\Phi_+)(D^4\Phi_+)^2-(D\Phi_+)(D^3\Phi_+)(D^5\Phi_+)\right]\Sigma\nonu\\
&&+\frac{5i}{4\omega}\cosh\Big(\frac{\Phi_+}{2}\Big)\left[(D\Phi_+)(D^4\Phi_+)(D^2\Phi_+)^2-4(D^4\Phi_+)(D^5\Phi_+)\right]\Sigma\\
&&+\frac{5}{\omega^2}\cosh^2\Big(\frac{\Phi_+}{2}\Big)(D\Phi_+)(D^3\Phi_+)(D^4\Phi_+)+\frac{5}{\omega^2}\cosh\Phi_+(D\Phi_+)(D^2\Phi_+)(D^5\Phi_+)\nonu\\
&&-\frac{30i}{\omega^5}\cosh\Big(\frac{\Phi_+}{2}\Big)\sinh(2\Phi_+) (D\Phi_+)(D^2\Phi_+)^2\Sigma\nonu \\&& -\frac{80i}{\omega^5}\cosh^3\Big(\frac{\Phi_+}{2}\Big)\cosh\Phi_+ (D^2\Phi_+)(D^3\Phi_+)\Sigma\nonu,\\[1cm]
 S_5 &=& -10 \left[(D^2\Phi_1)^2(D^6\Phi_1) - (D^2\Phi_2)^2(D^6\Phi_2)\right]\nonu\\
 &=& -\frac{5i}{\omega}\sinh\Big(\frac{\Phi_+}{2}\Big)\left[\frac{1}{4}(D^2\Phi_+)^4(D\Phi_+)+(D^2\Phi_+)^2(D^5\Phi_+)+2(D\Phi_+)(D^2\Phi_+)(D^6\Phi_+)\right]\Sigma\nonu\\
 && - \frac{5i}{\omega} \cosh\Big(\frac{\Phi_+}{2}\Big)\left[\frac{1}{2}(D\Phi_+)(D^2\Phi_+)^2(D^4\Phi_+)+(D^2\Phi_+)^3(D^3\Phi_+)\right]\Sigma\nonu\\
 &&-\frac{5}{\omega^2}\sinh\Phi_+ \left[2(D^2\Phi_+)^4 +4 (D^2\Phi_+)(D^6\Phi_+) +\frac{1}{2}(D^2\Phi_+)^2(D^3\Phi_+)(D\Phi_+) \right]\nonu\\
 &&-\frac{10}{\omega^2}\cosh\Phi_+ (D^2\Phi_+)^2(D^4\Phi_+) \nonu\\
 &&-\frac{40i}{\omega^5}\cosh\Big(\frac{\Phi_+}{2}\Big)\sinh^2\Phi_+\left[ (D^4\Phi_+) (D\Phi_+)+2 (D^2\Phi_+) (D^3\Phi_+)\right]\Sigma\nonu\\
 &&- \frac{20i}{\omega^5}\sinh\Big(\frac{\Phi_+}{2}\Big)\sinh^2\Phi_+\left[9(D^2\Phi_+)^2(D\Phi_+) +4 (D^5\Phi_+)\right]\Sigma\nonu\\
 && -\frac{80i}{\omega^5}\sinh\Big(\frac{\Phi_+}{2}\Big)\sinh(2\Phi_+) (D\Phi_+)(D^4\Phi_+)\Sigma\nonu\\
 &&-\frac{40}{\omega^6}\sinh^3\Phi_+ \left[4(D^2\Phi_+)^2+(D^3\Phi_+)(D\Phi_+)\right]  -\frac{160}{\omega^6}\cosh\Phi_+\sinh^2\Phi_+ (D^4\Phi_+),
\er
\br 
 S_6 &=& -\frac{5i}{\omega}\sinh\Big(\frac{\Phi_+}{2}\Big)\left[(D\Phi_+)(D^4\Phi_+)^2+2(D^2\Phi_+)(D^3\Phi_+)(D^4\Phi_+) \right]\Sigma \nonu \\ 
 && -\frac{5i}{\omega} \cosh\Big(\frac{\Phi_+}{2}\Big)(D\Phi_+)(D^2\Phi_+)^2(D^4\Phi_+)\Sigma\nonu\\
 && -\frac{10}{\omega^2}\sinh\Phi_+ (D^4\Phi_+)^2 -\frac{20}{\omega^2}\cosh\Phi_+ (D^2\Phi_+)^2(D^4\Phi_+)\nonu\\
 &&-\frac{80i}{\omega^5}\sinh\Big(\frac{\Phi_+}{2}\Big)\sinh(2\Phi_+)(D^2\Phi_+)(D^3\Phi_+)\Sigma\nonu\\
 && -\frac{80i}{\omega^5}\sinh\Big(\frac{3\Phi_+}{2}\Big)\cosh\Phi_+  (D^2\Phi_+)^2 (D\Phi_+)\Sigma\nonu\\
 &&-\frac{160}{\omega^6}\cosh^2\Phi_+\sinh\Phi_+  (D^2\Phi_+)^2,\\[0.8cm]
S_7 &=& -20 \left[ (D\Phi_1)(D^3\Phi_1)(D^2\Phi_1)^2 -  (D\Phi_2)(D^3\Phi_2)(D^2\Phi_2)^2\right]\nonu\\
&=& \frac{5i}{\omega}\cosh\Big(\frac{\Phi_+}{2}\Big)(D^3\Phi_+)(D^2\Phi_+)^3\Sigma -\frac{5i}{2\omega}\sinh\Big(\frac{\Phi_+}{2}\Big)(D\Phi_+)(D^2\Phi_+)^4\Sigma \nonumber\\
&& -\frac{15}{\omega^2}\sinh\Phi_+ (D\Phi_+)(D^2\Phi_+)^2(D^3\Phi_+)\nonu\\
&& + \frac{240i}{\omega^5}\cosh\Big(\frac{\Phi_+}{2}\Big)\sinh^2\Phi_+(D^2\Phi_+)(D^3\Phi_+)\Sigma\nonu\\
&& +\frac{240 i}{\omega^5}\cosh\Big(\frac{\Phi_+}{2}\Big)\sinh\Phi_+(D\Phi_+)(D^2\Phi_+)^2\Sigma\nonu\\
&&-\frac{80}{\omega^6}\sinh^3\Phi_+(D\Phi_+)(D^3\Phi_+)\nonu\\
&&+\frac{1280i}{\omega^9}\cosh^3\Big(\frac{\Phi_+}{2}\Big)\sinh^3\Phi_+(D\Phi_+)\Sigma,\\[0.8cm]
S_8 &=& 6\left[ (D^2\Phi_1)^5-(D^2\Phi_2)^5\]\nonu\\
&=& \frac{15i}{4\omega}\sinh\Big(\frac{\Phi_+}{2}\Big)(D\Phi_+)(D^2\Phi_+)^4\Sigma + \frac{15}{2\omega^2}\sinh\Phi_+ (D^2\Phi_+)^4 \nonu\\
&&+\frac{360i}{\omega^5}\sinh\Big(\frac{\Phi_+}{2}\Big)\sinh^2\Phi_+(D\Phi_+)(D^2\Phi_+)^2\Sigma+\frac{240}{\omega^6}\sinh^3\Phi_+ (D^2\Phi_+)^2\nonu \\
&& +\frac{960i}{\omega^9}\sinh\Big(\frac{\Phi_+}{2}\Big)\sinh^4\Phi_+ (D\Phi_+)\Sigma + \frac{384}{\omega^{10}}\sinh^5\Phi_+.
\er
Now, by summing up all the above results we find the following expression,
\br
 D_{t_5} \Phi_- &=&  \frac{i \t_0}{\omega}\left[(D\Phi_+)(D^8\Phi_+)-(D^2\Phi_+)(D^7\Phi_+)-(D\Phi_+)(D^2\Phi_+)^2(D^4\Phi_+)\right. \nonumber \\  && \left. \qquad  \,\,\, +(D^2\Phi_+)^3(D^3\Phi_+) - (D^3\Phi_+)(D^6\Phi_+)+(D^4\Phi_+)(D^5\Phi_+)  \right] \Sigma\nonumber\\
 &&+ \frac{i\t_1}{\omega}\left[\frac{3}{4}(D^2\Phi_+)^4(D\Phi_+) -2 (D^2\Phi_+)^2(D^5\Phi_+) -3(D\Phi_+)(D^2\Phi_+)(D^6\Phi_+) \right. \nonumber \\&&\left.\qquad\quad  -4 (D^2\Phi_+)(D^3\Phi_+)(D^4\Phi_+)-(D\Phi_+)(D^4\Phi_+)^2+2(D^9\Phi_+)\right] \Sigma \nonumber\\
 &&+\frac{\t_2}{\omega^2}\left[2 (D\Phi_+)(D^7\Phi_+)-4(D^2\Phi_+)(D^6\Phi_+)-2(D^3\Phi_+)(D^5\Phi_+)+2(D^4\Phi_+)^2   \nonumber \right. \\&& \qquad\,\,\, \left. +\frac{3}{2} (D^2\Phi_+)^4 -4(D\Phi_+)(D^2\Phi_+)^2(D^3\Phi_+)\right]\nonumber \\
 &&+\frac{\t_3}{\omega^2} \left[2(D\Phi_+)(D^2\Phi_+)(D^5\Phi_+)+4(D\Phi_+)(D^3\Phi_+)(D^4\Phi_+) -6(D^2\Phi_+)^2(D^4\Phi_+)\right. \nonu \\ &&\qquad \,\,\, \left.+4(D^8\Phi_+)  \right] \nonumber\\ 
 && +\frac{i}{\omega^5} \left[\t_4(D\Phi_+) (D^2\Phi_+)^2 +\t_5 (D\Phi_+)(D^4\Phi_+)+\t_6 (D^2\Phi_+)(D^3\Phi_+)+\t_7 (D^5\Phi_+) \right]\Sigma\nonumber\\
 &&+\frac{1}{\omega^6}\left[\t_8(D\Phi_+)(D^3\Phi_+) + \t_9(D^2\Phi_+)^2 +\t_{10}(D^4\Phi_+) \right] \nonumber\\
 && +\frac{i\t_{11} }{\omega^9} (D\Phi_+) \Sigma + \frac{\t_{12}}{\omega^{10}}, \label{eqD10}
\er 
where the $\t_k$ functions, with $k=1,..., 12$, are given as follows,
\br
\t_0 &=& \cosh\Big(\frac{\Phi_+}{2}\Big),\\
\t_1&=& \sinh\Big(\frac{\Phi_+}{2}\Big),\\
\t_2&=&\sinh\Phi_+,\\
\t_3 &=& \cosh\Phi_+,\\
 \t_4 &=& -5\sinh\Big(\frac{\Phi_+}{2}\Big)\left[13+12\cosh\Phi_+ +15\cosh2\Phi_+\right]\\
 \t_5 &=& 20\left[\cosh\Big(\frac{\Phi_+}{2}\Big) -\cosh\Big(\frac{3\Phi_+}{2}\Big)-4\cosh\Big(\frac{5\Phi_+}{2}\Big)\right]\\
 \t_6 &=& -5 \left[14\cosh\Big(\frac{\Phi_+}{2}\Big)-5\cosh\Big(\frac{3\Phi_+}{2}\Big)+7\cosh\Big(\frac{5\Phi_+}{2}\Big)\right]\\
 \t_7 &=& 40\cosh\Big(\frac{\Phi_+}{2}\Big)\sinh\Phi_+\big(1-3\cosh\Phi_+\big),\\
 \t_8 &=& -40 \sinh^3\Phi_+,\\
 \t_9 &=& -20 \big(5\sinh\Phi_+ +\sinh3\Phi_+\big),\\
 \t_{10} &=& -80\sinh\Phi_+\sinh2\Phi_+,\\
 \t_{11} &=& -160 \cosh\Big(\frac{\Phi_+}{2}\Big)\sinh^2\Phi_+\Big(2\sinh\Phi_+ - 5\sinh2\Phi_+\Big),\\
 \t_{12}&=& 384 \sinh^5 \Phi_+.
\er
%
%
\newpage
In order to derive the remaining equation we follow the same procedure as before for the smKdV equation, by imposing consistency of eqs. (\ref{asb1}) and (\ref{eqD10}), namely $D_{t_5}D\Phi_- = DD_{t_5} \Phi_-$, we get
\br
 D_{t_5}\Sigma &=& \frac{i\s_0}{\omega} \left[3(D\Phi_+)(D^2\Phi_+)(D^6\Phi_+)-\frac{3}{4}(D^2\Phi_+)^4(D\Phi_+) +2 (D^2\Phi_+)^2(D^5\Phi_+) \right. \nonumber \\&&\left.\qquad  +4 (D^2\Phi_+)(D^3\Phi_+)(D^4\Phi_+)+(D\Phi_+)(D^4\Phi_+)^2-2(D^9\Phi_+)\right]\nonumber\\
 &&+\frac{i\s_1}{\omega}\left[(D\Phi_+)(D^8\Phi_+)-(D^2\Phi_+)(D^7\Phi_+)-(D\Phi_+)(D^2\Phi_+)^2(D^4\Phi_+)\right. \nonumber \\&&\left.\qquad\,\,\, +(D^2\Phi_+)^3(D^3\Phi_+)  - (D^3\Phi_+)(D^6\Phi_+)+(D^4\Phi_+)(D^5\Phi_+)  \right] \nonu \\
 &&+\frac{1}{\omega^4} \Big[ \s_2 (D\Phi_+)(D^2\Phi_+)(D^3\Phi_+) +\s_3 (D^2\Phi_+)^3 +\s_4 (D^2\Phi_+)(D^4\Phi_+)  +\s_5 (D^6\Phi_+)  \Big] \Sigma\nonumber\\
 && +\frac{i}{\omega^5} \Big[ \s_6 (D\Phi_+)(D^2\Phi_+)^2 + \s_7 (D\Phi_+)(D^4\Phi_+) + \s_8 (D^2\Phi_+)(D^3\Phi_+) +\s_9 (D^5\Phi_+) \Big] \nonumber\\
 && +\frac{\s_{10}}{\omega^8} (D^2\Phi_+)\Sigma 
 +\frac{i \s_{11}}{\omega^{9}}(D\Phi_+),
\er
where the $\s_k$ functions are given by the following expressions,
\br
 \s_0 &=& \frac{1}{2}\cosh\Big(\frac{\Phi_+}{2}\Big),\\
 \s_1 &=& -\frac{1}{2}\sinh\Big(\frac{\Phi_+}{2}\Big),\\
 \s_2 &=& -20 \sinh^2\Big(\frac{\Phi_+}{2}\Big)\sinh\Phi_+, \\
 \s_3 &=& -15\cosh^2\Big(\frac{\Phi_+}{2}\Big)\sinh\Phi_+,\\
 \s_4 &=& 10\cosh^2\Big(\frac{\Phi_+}{2}\Big)\(3-7\cosh\Phi_+\),\\
 \s_5 &=& -20\cosh^2\Big(\frac{\Phi_+}{2}\Big)\sinh\Phi_+,\\
 \s_6 &=& \frac{5}{2}\cosh\Big(\frac{\Phi_+}{2}\Big)\(7+9\cosh\Phi_+\),\\
 \s_7 &=& -30 \cosh\Big(\frac{\Phi_+}{2}\Big)\sinh2\Phi_+,\\
 \s_8 &=& -60\cosh^3\Big(\frac{\Phi_+}{2}\Big)\sinh\Phi_+,\\
 \s_9 &=& -20\cosh\Big(\frac{\Phi_+}{2}\Big)\sinh^2\Phi_+,\\
 \s_{10} &=& 320\cosh^2\Big(\frac{\Phi_+}{2}\Big)\sinh^3\Phi_+,\\ 
 \s_{11} &=& -160 \cosh\Big(\frac{\Phi_+}{2}\Big)\sinh^4\Phi_+.
\er


\section{Conservation of mKdV momentum defect with respect to $ t_5 $}\label{appF}

Here we verify the conservation of the momentum of the mKdV with respect to $ t_5 $, namely
\br
\frac{dP}{dt_5}&=&\left[\, \frac{5}{8}(\pt_x\phi_1)^6+\frac{1}{16}(\pt^3_x\phi_1)^2-\frac{1}{8}\pt^2_x\phi_1\,\pt^4_x\phi_1+\frac{1}{8}\pt_x\phi_1\,\pt^5_x\phi_1-\frac{5}{8}(\pt_x\phi_1)^2(\pt^2_x\phi_1)^2\right.\nonumber\\&&\left. -\frac{5}{4}(\pt_x\phi_1)^3\pt^3_x\phi_1 + i\bp_1\pt_x\bp_1\left(\frac{35}{16}\pt_x\phi_1\,\pt^3_x\phi_1+\frac{5}{8}(\pt^2_x\phi_1)^2-\frac{25}{8}(\pt_x\phi_1)^4\right)+\frac{i}{8}\pt_x\bp_1\pt^4_x\bp_1\right.\nonumber\\&&\left. -\frac{i}{8}\pt^2_x\bp_1\pt^3_x\bp_1- \frac{i}{16}\bp_1\pt^5_x\bp_1-\frac{5i}{8}(\pt_x\phi_1)^2\pt_x\bp_1\pt^2_x\bp_1+\frac{15i}{16}\pt_x\phi_1\,\pt^2_x\phi_1\bp_1\pt^2_x\bp_1 \right. \nonu\\&&\left. +\frac{15i}{16}(\pt_x\phi_1)^2\bp_1\pt^3_x\bp_1\, \right]_{x=0}-\left[\phi_1\rightarrow \phi_2,\, \bp_1\rightarrow\bp_2\right]_{x=0}.
\er
Using the equations of motion \eqref{t5u} and \eqref{t5psi}, we can write the above expression in the following form,
\br
\frac{dP}{dt_5}&=&\left[\, 2\pt_x\phi_1\,\pt_{t_5}\phi_1-\frac{1}{8}(\pt_x\phi_1)^6+\frac{1}{16}(\pt^3_x\phi_1)^2+\frac{5}{8}(\pt_x\phi_1)^2(\pt^2_x\phi_1)^2-\frac{1}{8}\pt_x\phi_1\,\pt^4_x\phi_1\right.\nonumber\\
&&\left.-\frac{5i}{8}(\pt_x\phi_1)^2\pt_x\bp_1\pt^2_x\bp_1-\frac{5i}{8}\pt_x\phi_1\,\pt^2_x\phi_1\bp_1\pt^2_x\bp_1+\frac{5i}{8}\pt_x\phi_1\,\pt^3_x\phi_1\bp_1\pt_x\bp_1-i\bp_1\pt_{t_5}\bp_1\right.\nonumber\\
&&\left.+\frac{i}{8}\pt_x\bp_1\pt^4_x\bp_1-\frac{i}{8}\pt^2_x\bp_1\pt^3_x\bp_1\, \right]_{x=0}-\left[\phi_1\rightarrow \phi_2,\, \bp_1\rightarrow\bp_2\right]_{x=0}.
\er
In terms of the variables $\phi_{\pm}=\phi_1\pm\phi_2$ and $\bp_{\pm}=\bp_1\pm\bp_2$, it reads
\br
\frac{dP}{dt_5}&=&\left[\, \pt_x\phi_-\,\pt_{t_5}\phi_++\pt_x\phi_+\,\pt_{t_5}\phi_-+\frac{1}{16}\pt^3_x\phi_-\,\pt^3_x\phi_+-\frac{3}{128}(\pt_x\phi_-)^5\pt_x\phi_+-\frac{3}{128}(\pt_x\phi_+)^5\pt_x\phi_-\right.\nonumber\\
&&\left.-\frac{1}{16}\pt^2_x\phi_+\,\pt^4_x\phi_--\frac{1}{16}\pt^2_x\phi_-\,\pt^4_x\phi_+-\frac{5}{64}(\pt_x\phi_-)^3(\pt_x\phi_+)^3+\frac{5}{32}(\pt_x\phi_-)^2\pt^2_x\phi_-\,\pt^2_x\phi_+\right.\nonu\\
&&\left. +\frac{5}{32}\pt_x\phi_-\,\pt_x\phi_+(\pt^2_x\phi_-)^2+\frac{5}{32}\pt_x\phi_-\,\pt_x\phi_+(\pt^2_x\phi_+)^2+\frac{i}{16}\pt_x\bp_-\pt^4_x\bp_+ +\frac{i}{16}\pt_x\bp_+\pt^4_x\bp_-\right.\nonu\\
&&\left. -\frac{5i}{32}\pt_x\phi_-\,\pt_x\phi_+\left(\pt_x\bp_-\pt^2_x\bp_-+\pt_x\bp_+\pt^2_x\bp_+\right)-\frac{i}{16}\pt^2_x\bp_-\pt^3_x\bp_+-\frac{i}{16}\pt^2_x\bp_+\pt^3_x\bp_- \right. \nonu\\
&&\left.-\frac{5i}{64}\left(\bp_-\pt^2_x\bp_-+\bp_+\pt^2_x\bp_+\right)\left(\pt_x\phi_+\,\pt^2_x\phi_-+\pt_x\phi_-\,\pt^2_x\phi_+\right)-\frac{i}{2}\bp_-\pt_{t_5}\bp_+-\frac{i}{2}\bp_+\pt_{t_5}\bp_-\right.\nonu\\&&
\left.+\frac{5i}{64}\left(\bp_-\pt_x\bp_-+\bp_+\pt_x\bp_+\right)\left(\pt_x\phi_+\,\pt^3_x\phi_-+\pt_x\phi_-\,\pt^3_x\phi_+\right)\right.\nonu\\&&
\left. +\frac{5i}{64}\left(\bp_-\pt_x\bp_++\bp_+\pt_x\bp_-\right)\left(\pt_x\phi_-\,\pt^3_x\phi_-+\pt_x\phi_+\,\pt^3_x\phi_+\right)\right.\nonu\\&&
\left.+\frac{5i}{64}\left(\bp_-\pt^2_x\bp_++\bp_+\pt^2_x\bp_-\right)\left(\pt_x\phi_-\,\pt^2_x\phi_-+\pt_x\phi_+\,\pt^2_x\phi_+\right)\right.\nonu\\&&
\left.-\frac{5i}{64}\left(\pt_x\bp_-\pt^2_x\bp_++\pt_x\bp_+\pt^2_x\bp_-\right)\left((\pt_x\phi_-)^2+(\pt_x\phi_+)^2\right)\, \right]_{x=0}.
\er
Now using the B\"acklund equations (\ref{dxphi-})--(\ref{dxf1}), (\ref{btt5phi}), (\ref{btt5psi}), we find
\br
\frac{dP}{dt_5}&=&\left[-\frac{i}{\o}\left(2\cosh\left(\frac{\phi_+}{2}\right)f_1\pt_{t_5}\bp_++\sinh\left(\frac{\phi_+}{2}\right)\pt_{t_5}\phi_+f_1\bp_+\right)+\frac{4}{\o^2}\sinh\phi_+\pt_{t_5}\phi_+
\right.\nonumber\\
&&\left.
+\frac{i}{8\o^2}\left(\cosh^2\left(\frac{\phi_+}{2}\right)\bp_+\pt^4_x\bp_+-\frac{1}{4}\sinh\phi_+\pt_x\phi_+\bp_+\pt^3_x\bp_+\right)\right.\nonumber\\&&\left.+\frac{i}{16\o^2}\bp_+\pt_x\bp_+\left(\frac{1}{2}\sinh\phi_+\left((\pt_x\phi_+)^3-\pt^3_x\phi_+\right)-4\cosh^2\left(\frac{\phi_+}{2}\right)\pt_x\phi_+\pt^2_x\phi_+\right)\right.\nonumber\\&&\left.
+\frac{i}{8\o^2}\bp_+\pt^2_x\bp_+\left(\frac{1}{4}\sinh\phi_+\pt^2_x\phi_+-\cosh^2\left(\frac{\phi_+}{2}\right)(\pt_x\phi_+)^2\right)
\right.\nonumber\\&&\left.
+\frac{5i}{8\o^5}\cosh^3\Big(\frac{\phi_+}{2}\Big)f_1\bp_+\left(\sinh\phi_+\left(3(\pt_x\phi_+)^3+4\pt^3_x\phi_+\right)+2\pt_x\phi_+\pt^2_x\phi_+(7\cosh\phi_+-3)\right)
\right.\nonumber\\&&\left.-
\frac{5i}{2\o^6}\cosh^2\left(\frac{\phi_+}{2}\right)\sinh\phi_+\left(\sinh\phi_+\bp_+\pt^2_x\bp_++3\cosh^2\left(\frac{\phi_+}{2}\right)\pt_x\phi_+\bp_+\pt_x\bp_+\right)
\right.\nonumber\\&&\left. -
\frac{40i}{\o^9}\cosh^3\left(\frac{\phi_+}{2}\right)\sinh^3\phi_+\pt_x\phi_+f_1\bp_+\,\right]_{x=0}.
\er
Note that the first term can be written as  follows,
\begin{equation}
\pt_{t_5}\left(2\cosh\left(\frac{\phi_+}{2}\right)f_1\bp_+\right)-2\cosh\left(\frac{\phi_+}{2}\right)\pt_{t_5}f_1\bp_+,
\end{equation}
and then,
\br\label{dt5P}
\frac{dP}{dt_5}&=&\left[\pt_{t_5}\left(\frac{4}{\o^2}\cosh\phi_+-\frac{2i}{\o}\cosh\left(\frac{\phi_+}{2}\right)f_1\bp_+\right)+\frac{2i}{\o}\cosh\left(\frac{\phi_+}{2}\right)\pt_{t_5}f_1\bp_+
\right.\nonumber\\
&&\left.+
\frac{i}{8\o^2}\left(\cosh^2\left(\frac{\phi_+}{2}\right)\bp_+\pt^4_x\bp_+-\frac{1}{4}\sinh\phi_+\pt_x\phi_+\bp_+\pt^3_x\bp_+\right)\right.\nonumber\\&&\left.+\frac{i}{16\o^2}\bp_+\pt_x\bp_+\left(\frac{1}{2}\sinh\phi_+\left((\pt_x\phi_+)^3-\pt^3_x\phi_+\right)-4\cosh^2\left(\frac{\phi_+}{2}\right)\pt_x\phi_+\pt^2_x\phi_+\right)\right.\nonumber\\&&\left.
+\frac{i}{8\o^2}\bp_+\pt^2_x\bp_+\left(\frac{1}{4}\sinh\phi_+\pt^2_x\phi_+-\cosh^2\left(\frac{\phi_+}{2}\right)(\pt_x\phi_+)^2\right)
\right.\nonumber\\&&\left. +
\frac{5i}{8\o^5}\cosh^3\Big(\frac{\phi_+}{2}\Big)f_1\bp_+\!\left(\sinh\phi_+\left(3(\pt_x\phi_+)^3+4\pt^3_x\phi_+\right)+2\pt_x\phi_+\pt^2_x\phi_+(7\cosh\phi_+-3)\right)
\right.\nonumber\\&&\left. -
\frac{5i}{2\o^6}\cosh^2\left(\frac{\phi_+}{2}\right)\sinh\phi_+\left(\sinh\phi_+\bp_+\pt^2_x\bp_++3\cosh^2\left(\frac{\phi_+}{2}\right)\pt_x\phi_+\bp_+\pt_x\bp_+\right)
\right.\nonumber\\&&\left.-
\frac{40i}{\o^9}\cosh^3\left(\frac{\phi_+}{2}\right)\sinh^3\phi_+\pt_x\phi_+f_1\bp_+\,\right]_{x=0}.
\er
By using eq. (\ref{btt5psi}) we can write eq (\ref{dt5P}) as a total time ($t_3$) derivative, and finally we obtain that
\begin{equation}
\mathcal{P}=P-\Big[\frac{4}{\o^2}\cosh\phi_+-\frac{2i}{\o}\cosh\left(\frac{\phi_+}{2}\right)f_1\bp_+\Big]_{x=0},
\end{equation}
is the modified conserved momentum, which includes the same defect contribution previously derived by applying the $t_3$ derivative.


\vskip 2cm
\noindent
{\bf Acknowledgements} \\
\vskip .1cm \noindent
{ALR would like to thank to  FAPESP S\~ao Paulo Research Foundation for financial support under the PhD fellowship 2015/00025-9. NIS thanks CNPq for financial support. AHZ and JFG thank CNPq and FAPESP for support. The authors thank the referee for useful and interesting comments.  
}


\end{document}